\documentclass[10pt,a4paper,twocolumn]{IEEEtran}

\usepackage{amssymb}
\usepackage{graphicx}
\usepackage{multirow}
\usepackage{float}
\usepackage{amsmath,amssymb,amsfonts,mathrsfs}%,amsthm
\usepackage{color}
\usepackage{subfig}

\newtheorem{theorem}{\textbf{Theorem}}
\newtheorem{lemma}{\textbf{Lemma}}
\newtheorem{corollary}{\textbf{Corollary}}

\begin{document}
\renewcommand{\thepage}{14--\arabic{page}}
\setcounter{page}{1}
\setlength{\topsep}{0.0cm}
\setlength{\itemsep}{0.0em}

\title{\Large\bf Regularity/Controllability/Observability of an NDS with Descriptor Form Subsystems and Generalized LFTs} % Title, preferably not more than 10 words.

\author{Tong Zhou % <-this % stops a space
\thanks{This work was supported in part by the NNSFC under Grant 61733008, 5191101838 and 61573209.}% <-this % stops a space
\thanks{Tong Zhou is with the Department of Automation and BNRist, Tsinghua University, Beijing, 100084, P.~R.~China
        {(email: {\tt\small tzhou@mail.tsinghua.edu.cn}).}}%
}
\maketitle

\begin{abstract}                          % Abstract of not more than 200 words.
This paper investigates regularity, controllability and observability for a networked dynamic system (NDS) with its subsystems being described in a descriptor form and system matrices of each subsystem being represented by a generalized linear fractional transformation (GLFT) of its parameters. Except a well-posedness condition, no any other constraints are put on either parameters or connections of a subsystem. Based on the Kronecker canonical form (KCF) of a matrix pencil, some matrix rank based necessary and sufficient conditions are established respectively for the regularity and complete controllability/observability of the NDS, in which the associated matrix depends affinely on both subsystem parameters and subsystem connections. These conditions keep the property that all the involved numerical computations are performed on each subsystem independently, which is attractive in the analysis and synthesis of a large scale NDS. Moreover, some explicit and easily checkable requirements are derived for subsystem dynamics/parameters with which a completely  controllable/observable NDS can be constructed more easily.
\end{abstract}


\setlength{\itemsep}{0.0em}

\section{Introduction}
\setlength{\itemsep}{0.0em}

Controllability and observability are extensively regarded as essential requirements for a system to work properly. Various system performances, such as minimizing the tracking error of a servo system, the existence of an optimal control measured by a $H_{2}$/$H_{\infty}$ norm, possibilities of stabilizing a plant and/or locating its poles to a prescribed desirable area, convergence of a state estimation procedure like the widely adopted  Kalman filter, etc., usually ask that the plant at hand is controllable  and/or observable (\cite{ksh2000,sbkkmpr2011,zdg1996,zyl2018}). On the other hand, the associated concepts about system controllability and observability have been well developed and studied, and various criteria have been obtained for different types of systems under distinguished working situations. Among them, the most widely adopted ones appear to be the PBH test, the full rank condition of the controllability/observability matrix, various measures for the controllability/observability Gramian of a system, etc. (\cite{ksh2000,zdg1996,zyl2018}).

It has now been made clear that for many systems, both controllability and observability are a generic system property. That is, rather than a numerical value of the system matrices, it is the connections among the system states, inputs and outputs that determine its controllability/observability (\cite{dcv2003,Lin1974,zyl2018}). \cite{ccvfb2012} clarifies that in order to guarantee the structural controllability/observability of a networked dynamic system (NDS), rather than its in/out-degree distributions, it is the dynamics of its subsystems that determine the minimal input/output number. Some distributed procedures are developed in \cite{cpakj2017} for the verification of the structural controllability of several types of NDSs consisting of small subsystems. Graph theories and Cartesian products have been respectively adopted in \cite{np2013} and \cite{cam2014} in the studies of NDS controllability/observability. On the other hand, when the state transition matrix is known, it is proved in \cite{Olshevsky2014} that the problem of finding the sparest input/output matrix such that the associated NDS is controllable/observable, is NP-hard. Some measures are suggested in \cite{pzb2014} to quantitatively analyze difficulties of controlling a system. \cite{Zhou2018} makes it clear that in order to construct a controllable/observable NDS, each subsystem must be controllable/observable. It has also been found there that the minimal input/output number of each subsystem is equal to the maximum geometric multiplicity of its state transition matrix.

While various results have been obtained for controllability/observability of an NDS, many theoretical issues still require further efforts, which include influences from subsystem dynamics, subsystem connections, etc., to the controllability/observability of the whole system. Another challenging issue is computational costs and numerical stability (\cite{cam2014,cpakj2017,zyl2018}).

On the other hand, in designing a system, it is quite important to at first construct a plant that has the capability to achieve the expected performances. When an NDS is to be designed, this problem include subsystem dynamics/parameter selections and designs of subsystem connections. To achieve this objective, it appears helpful to establish an explicit dependence of the system achievable performances on  its subsystem dynamics and parameters, as well as on its subsystem connections (\cite{Kailath1980,sbkkmpr2011,zdg1996}).

In the description of plant dynamics, descriptor systems have proved to be an appropriate model. It is widely believed that compared with the extensively adopted state space model, a descriptor system is more suitable in keeping structural information of plant dynamics. Like the state space model, this model has also been adopted in many fields including engineering, economy, biology, etc., and has attracted extensive research attentions (\cite{Dai1989, Duan2010, Kailath1980}). In addition, it has been argued in \cite{hv2004} that a generalized linear fractional transformation (GLFT) is more efficient in describing dependence of system matrices on its parameters, in the sense that the associated matrices have a lower dimension.

In this paper, we investigate regularity and controllability/observability for NDSs in which each subsystem is described by a descriptor form like model, while its system matrices are described by a GLFT. Results of \cite{Zhou2015,zyl2018,Zhou2019} have been extended, in which each subsystem is described by a state space like model and for each subsystem, all of its system matrices are assumed to be known or to be described by a linear fractional transformation (LFT). In this investigation, except that the NDS is required to be well-posed, including the whole system and each of its subsystems, there are neither any other restrictions on a subsystem first principle parameter (FPP), nor any other restrictions on an element of the subsystem connection matrix (SCM). Using the Kronecker canonical form (KCF) of a matrix pencil, several rank based conditions are established respectively for the regularity of the NDS and its complete controllability/observability. In these conditions, the associated matrix affinely depends on both the SCM and a matrix formed by the parameters of each subsystem, which agrees well with that of \cite{Zhou2015,zyl2018,Zhou2019}. These conditions keep the properties of the results reported in \cite{Zhou2015,zyl2018,Zhou2019} that in obtaining the associated matrices, all the required  numerical computations are performed on each subsystem independently. This is quite attractive, as it means that the associated condition verification is scalable for an NDS constructed from a large number of subsystems. In addition, this explicit relation between the condition related matrix and subsystem parameters/connections may be helpful in system topology designs and subsystem dynamics/parameter selections. Moreover, some explicit conditions are obtained for both the dynamics and parameters of a subsystem with which a completely controllable/observable NDS can be constructed more easily. These conditions can be verified for each subsystem separately, which makes them convenient in the design of a large scale NDS.

Compared with other models for an NDS, such as the well known Roesser model and FM models, etc. (\cite{fm1978,np2013,Roesser1975,sbkkmpr2011,Zhou2006}), the model adopted in this paper allows not only distinctive subsystem dynamics, but also arbitrary subsystem connections. The obtained results, however, are also applicable to these NDS models, while computational costs will be significantly reduced further. In addition, these results establish simple relations between NDS regularity/controllability/observability and its subsystem parameters/connections.

The outline of this paper is as follows. At first, in Section 2, a descriptor form like model is given for an NDS, together with some preliminary results. Regularity, controllability and observability of an NDS are investigated in Section 3 under the assumption that the parameters of each subsystem are known. These results are extended in Section 4 to an NDS, in which system matrices of each subsystem depends on its parameters through a GLFT. Section 5 provides an artificial example to illustrate the theoretical results. Finally, some concluding remarks are given in Section 6 in which several further issues are discussed. An appendix is included to give proofs of some technical results.

The following notation and symbols are adopted. $\cal C$ and ${\cal R}^{n}$ stand respectively for the set of complex numbers and the $n$ dimensional real Euclidean space. ${\rm\bf det} \left(\cdot\right)$ represents the determinant of a square matrix, ${\rm\bf null} \left(\cdot\right)$ and ${\rm\bf span} \left(\cdot\right)$ the (right) null space of a matrix and the space spanned by the columns of a matrix, while $\cdot^{\perp}$ the matrix whose columns form a base of the (right) null space of a matrix. ${\rm\bf diag}\{X_{i}|_{i=1}^{L}\}$ denotes a block diagonal matrix with its $i$-th diagonal block being $X_{i}$, while ${\rm\bf col}\{X_{i}|_{i=1}^{L}\}$ the vector/matrix stacked by $X_{i}|_{i=1}^{L}$ with its $i$-th row block vector/matrix being $X_{i}$. $I_{m}$, $0_{m}$ and $0_{m\times n}$ represent respectively the $m$ dimensional identity matrix, the $m$ dimensional zero column vector and the $m\times n$ dimensional zero matrix. The subscript is usually omitted if it does not lead to confusions. The superscripts $T$ and $H$ are used to denote respectively the transpose and the conjugate transpose of a matrix/vector. A matrix valued function (MVF) is said to be of full normal column/row rank (FNCR/FNRR), if there exists a value for its variable(s), at which the value of the MVF is of full column/row rank (FCR/FRR).

\section{System Description and Some Preliminaries}

In many actual engineering problems, an NDS is constructed from subsystems with distinguished input-output relations. A possible approach to describe the dynamics of a general linear time invariant (LTI) NDS is to represent the dynamics of each subsystem using an ordinary model, divide the inputs/outputs of each subsystem into external and internal ones, express subsystem interactions through connecting internal outputs of a subsystem to internal inputs of some other subsystems. This approach has been adopted in \cite{Zhou2015,zyl2018,Zhou2019} in which the dynamics of each subsystem is described by a state space model. To reflect structure information of each subsystem in an NDS more appropriately, a descriptor form is adopted in this paper for its dynamics description. In addition, in order to express the dependence of the system matrices of a subsystem on its FPPs, it is assumed that each of their elements is a function of these parameters. More precisely, the following model is used in this paper to describe the dynamics of the $i$-th subsystem ${\bf{\Sigma}}_i$ with $1\leq i\leq N$ in an NDS $\bf \Sigma$ that is composed from $N$ subsystems,
\begin{eqnarray}
& &\!\!\!\! \left[\! \begin{array}{c}
{E(i,p_{i})}{\delta({{x}}(t,i))}\\
{z(t,i)}\\
{{y}(t,i)}
\end{array} \!\right] \nonumber\\
&=& \!\!\!\! \left[\! \begin{array}{ccc}
{A_{\rm\bf xx}{(i,p_{i})}} & {A_{\rm\bf xv}{(i,p_{i})}} & {B_{\rm\bf x}{(i,p_{i})}}\\
{A_{\rm\bf zx}{(i,p_{i})}} & {A_{\rm\bf zv}{(i,p_{i})}} & {B_{\rm\bf z}{(i,p_{i})}}\\
{C_{\rm\bf x}{(i,p_{i})}} & {C_{\rm\bf v}{(i,p_{i})}} & {D_{\rm\bf u}{(i,p_{i})}}
\end{array} \!\right]\left[\! \begin{array}{c}
{x(t,i)}\\
{v(t,i)}\\
{u(t,i)}
\end{array} \!\right]
\label{eqn:1}
\end{eqnarray}
Here, $\delta(\cdot)$ represents either a forward time shift operation or the derivative of a function with respect to time. That is, the above model can be either continuous time or discrete time. Moreover, $p_{i}$ stands for the vector that consists of all the parameters in the subsystem ${\bf{\Sigma}}_i$, which may be the masses, spring and damper coefficients of a mechanical system, concentrations and reaction ratios  of a chemical/biological process, resistors, inductor and capacitor coefficients of an electronic/electrical system, etc. These parameters are usually called a FPP as they can be selected or adjusted in designing an actual system. In addition, $t$ denotes the temporal variable, $x(t,i)$ its state vector, $u(t,i)$ and $y(t,i)$ respectively its external input and output vectors,  $v(t,i)$ and $z(t,i)$ respectively its internal input and output vectors which denote signals received from other subsystems and signals transmitted to other subsystems.

To emphasize the simultaneous existence of both external and internal inputs/outputs in the above description, it is called a descriptor form like model throughout this paper.

Denote vectors ${\rm\bf col}\left\{v(t,i)|_{i=1}^{N}\right\}$ and ${\rm\bf col}\left\{z(t,i)|_{i=1}^{N}\right\}$ respectively by $v(t)$ and $z(t)$. It is assumed in this paper that the interactions among subsystems of the NDS $\rm\Sigma$ are described by
\begin{equation}
v(t)=\Phi z(t)
\label{eqn:2}
\end{equation}
The matrix $\Phi$ is called the subsystem connection matrix (SCM), which describes influences between different subsystems of an NDS and can also be a MVF of some FPPs. This dependence, however, is omitted for conciseness. A graph can be assigned to an NDS when each subsystem  is regarded as a node and each nonzero element in the SCM $\Phi$ as an edge. This graph is usually referred as the structure or topology of the associated NDS.

Compared with the subsystem model adopted in \cite{Zhou2015,zyl2018,Zhou2019}, it is clear that each of its system matrices in the above model, that is, ${A_{\rm\bf *\#}(i)}$, ${B_{\rm\bf *}(i)}$, ${C_{\rm\bf *}(i)}$ with ${\rm\bf *,\#}={\rm\bf x}$, ${\rm\bf u}$, ${\rm\bf v}$, ${\rm\bf y}$ or ${\rm\bf z}$, as well as the matrices $E(i)$ and ${D_{\rm\bf u}(i)}$, is a MVF of the parameter vector $p_{i}$. This reflects the fact that in an actual system, elements of its system matrices are usually not algebraically independent of each other, and some of them can even not be changed in system designs. It can therefore be declared that this model is more convenient in investigating influences of system parameters on the behaviors of a dynamic plant.

Obviously, the aforementioned model is also applicable to situations in which we are only interested in the influences from part of the subsystem FPPs on the performances of the whole NDS. This can be simply done through fixing all other FPPs to a particular numerical value.

The following assumptions are adopted throughout this paper for the NDS $\rm\bf\Sigma$.
\begin{itemize}
\item The dimensions of the vectors $u(t,i)$, $v(t,i)$, $x(t,i)$, $y(t,i)$ and $z(t,i)$ are respectively $m_{{\rm\bf u}i}$, $m_{{\rm\bf v}i}$, $m_{{\rm\bf x}i}$, $m_{{\rm\bf y}i}$ and $m_{{\rm\bf z}i}$.
\item Each subsystem ${\bf{\Sigma}}_i$, $i=1,2,\cdots,N$, is well posed.
\item The whole NDS ${\bf{\Sigma}}$ is well posed.
\end{itemize}

Note that the first assumption is only for indicating the size of the involved vectors. On the other hand, well-posedness is an essential requirement for a system to work properly (\cite{Kailath1980,zdg1996,zyl2018}). It appears safe to declare that all the above three assumptions must be satisfied for a practical system. Therefore, the adopted assumptions seem not very restrictive in actual applications.

Using these symbols, define integers $M_{{\rm\bf x}i}$, $M_{{\rm\bf
v}i}$, $M_{\rm\bf x}$ and $M_{\rm\bf v}$ as $M_{\rm\bf
x}={\sum_{k=1}^{N} m_{{\rm\bf x}k}}$, $ M_{\rm\bf v}={\sum_{k=1}^{N}
m_{{\rm\bf v}k}}$, and $M_{{\rm\bf x}i}=M_{{\rm\bf v}i}=0$ when
$i=1$, $M_{{\rm\bf x}i}={\sum_{k=1}^{i-1} m_{{\rm\bf x}k}}$,
$M_{{\rm\bf v}i}={\sum_{k=1}^{i-1} m_{{\rm\bf v}k}}$ when $2\leq
i\leq N$. Obviously, the SCM $\Phi$ is a $M_{\rm\bf v}\times M_{\rm\bf z}$ dimensional real matrix. These definitions are adopted throughout the rest of this paper.

The following results on a matrix pencil are required in deriving a computationally checkable necessary and sufficient condition for the regularity, controllability or observability of the aforementioned NDS, which can be found in many references, for example, \cite{bv1988,Kailath1980,it2017}.

For two arbitrary $m\times n$ dimensional real matrices $G$ and $H$, a first degree matrix valued polynomial (MVP) $\Psi(\lambda)=\lambda G+H$ is called a matrix pencil. When $m=n$ and ${\rm\bf det}(\Psi(\lambda))\not\equiv 0$, this matrix pencil is called regular. A regular matrix pencil is called strictly regular if the associated matrices $G$ and $H$ are both invertible. A matrix pencil $\bar{\Psi}(\lambda)$ is said to be strictly equivalent to the matrix pencil ${\Psi}(\lambda)$, if there exist two invertible real matrices $U$ and $V$ satisfying $\Psi(\lambda)=U\bar{\Psi}(\lambda)V$.

Given a positive integer $m$, two $m\times m$ dimensional matrix pencils $K_{m}(\lambda)$ and $N_{m}(\lambda)$, a $m\times (m+1)$ dimensional matrix pencil $L_{m}(\lambda)$, as well as a $(m+1)\times m$ dimensional matrix pencil $J_{m}(\lambda)$, are defined respectively as follows.
\begin{eqnarray}
& &
\hspace*{-1.0cm} K_{m}(\lambda)\!=\!\lambda I_{m}\!+\!\left[\!\!\begin{array}{cc}
0 & I_{m-1} \\ 0 & 0 \end{array}\!\!\!\right]\!\!,\hspace{0.1cm}
N_{m}(\lambda)\!=\!\lambda \!\left[\!\!\begin{array}{cc}
0 & I_{m-1} \\ 0 & 0 \end{array}\!\!\!\right] \!+\! I_{m} \label{eqn:3} \\
& &
\hspace*{-1.0cm} L_{m}(\lambda)=\left[\begin{array}{cc}
K_{m}(\lambda) & \left[\begin{array}{c} 0 \\ 1 \end{array}\right] \end{array}\right],\hspace{0.2cm}
J_{m}(\lambda)= \left[\begin{array}{c}
K_{m}^{T}(\lambda) \\ \left[0 \;\;\;\;\; 1\right] \end{array}\right] \label{eqn:4}
\end{eqnarray}

Obviously, $J_{m}(\lambda)=L_{m}^{T}(\lambda)$. In the following analysis for the regularity, controllability and observability of the NDS $\rm\bf\Sigma$, however, the roles of these two matrix pencils are completely different. To emphasize these differences, as well as to have a clear presentation, it appears better to adopt different symbols for them. Moreover, when $m=0$, $L_{m}(\lambda)$ is a $0\times 1$ zero matrix, while $J_{m}(\lambda)$ is a $1\times 0$ zero matrix.

From the definitions of the matrix pencils $K_{m}(\lambda)$, $N_{m}(\lambda)$, $L_{m}(\lambda)$ and $J_{m}(\lambda)$, the following characteristics can be straightforwardly established for their ranks and the associated null spaces. The details of the proof are omitted due to their obviousness, but can be found in \cite{Zhou2019}.

\begin{lemma} For any positive integer $m$, the matrix pencils defined in Equations (\ref{eqn:3}) and (\ref{eqn:4}) respectively have the following properties.  % \vspace{-0.25cm}
\begin{itemize}
\item An $m\times m$ dimensional strictly regular matrix pencil $H_{m}(\lambda)$ is rank deficient only at some isolated values of the complex variable $\lambda$ which are different from zero. Moreover, the number of these values is equal to $m$.
\item The matrix pencil $N_{m}(\lambda)$ is always of full rank (FR).
\item The matrix pencil $J_{m}(\lambda)$ is always of FCR.
\item The matrix pencil $K_{m}(\lambda)$ is singular only at $\lambda=0$. Moreover, ${\rm\bf Null}\left\{ K_{m}(0)\right\}={\rm\bf Span}\left({\rm\bf col}\left\{1, 0_{m-1}\right\}\right)$.
\item The matrix pencil $L_{m}(\lambda)$ is not of FCR at each $\lambda\in{\cal C}$. Moreover, ${\rm\bf Null}\left\{ L_{m}(\lambda)\right\}=
{\rm\bf Span}\left({\rm\bf col}\left\{\left. 1, (-\lambda)^{j}\right|_{j=1}^{m}\right\}\right)$.
\end{itemize}
\label{lemma:0}
\end{lemma}

It is well known that any matrix pencil is strictly equivalent to a block diagonal form with its diagonal blocks being strictly regular, or in the form of the matrix pencils $K_{\star}(\lambda)$, $N_{\star}(\lambda)$, $L_{\star}(\lambda)$ and $J_{\star}(\lambda)$, which is extensively called the Kronecker canonical form (KCF). More precisely, we have the following results (\cite{bv1988,Gantmacher1959,it2017}).

\begin{lemma}
For any two $m\times n$ dimensional real matrices $G$ and $H$, there exist some unique nonnegative integers $\xi_{H}$, $\zeta_{K}$, $\zeta_{L}$, $\zeta_{N}$, $\zeta_{J}$, $\xi_{L}(j)|_{j=1}^{\zeta_{L}}$ and
$\xi_{J}(j)|_{j=1}^{\zeta_{J}}$, some unique positive integers $\xi_{K}(j)|_{j=1}^{\zeta_{K}}$ and $\xi_{N}(j)|_{j=1}^{\zeta_{N}}$, as well as a strictly regular $\xi_{H}\times \xi_{H}$ dimensional matrix pencil $H_{\xi_{H}}(\lambda)$, such that the matrix pencil $\bar{\Psi}(\lambda)=\lambda G+H$ is strictly equivalent to a block diagonal form ${\Psi}(\lambda)$ with the following  definition,
\begin{eqnarray}
{\Psi}(\lambda)\!\!&=&\!\!{\rm\bf diag}\!\left\{H_{\xi_{H}}(\lambda),\;K_{\xi_{K}(j)}(\lambda)|_{j=1}^{\zeta_{K}},\; L_{\xi_{L}(j)}(\lambda)|_{j=1}^{\zeta_{L}}, \right.\nonumber \\
& & \hspace*{2.0cm}\left. N_{\xi_{N}(j)}(\lambda)|_{j=1}^{\zeta_{N}},\; J_{\xi_{J}(j)}(\lambda)|_{j=1}^{\zeta_{J}}\!\right\}
\label{eqn:5}
\end{eqnarray}
\label{lemma:1}
\end{lemma}

Descriptor systems are extensively utilized in describing input-output relations of a dynamic plant. It is believed that compared with a state space model, a descriptor system is more suitable in describing system constraints and keeping system structures (\cite{Dai1989,Duan2010,Kailath1980}). More precisely, if the input-output relations of an LTI plant can be described by the following equations,
\begin{equation}
E{\delta({{x}}(t))}=Ax(t)+Bu(t), \hspace{0.5cm} y(t)=Cx(t)+Du(t)
\label{eqn:6}
\end{equation}
in which $A$, $B$, $C$, $D$ and $E$ are constant real matrices with consistent dimensions, then this plant is called a descriptor system. It is said to be
regular if there exists a $\lambda\in{\cal C}$, such that ${\rm\bf det}(\lambda E - A)\neq 0$. When the initial states of a descriptor system can be uniquely determined by its outputs over the whole time interval, it is said to be completely observable. On the other hand, a descriptor system is said to be completely controllable, if for any two given state vectors $x_{0}$ and $x_{f}$, there exists a finite time $t_{f}$ and a series/sequence of control input vectors $u(t)|_{t=0}^{t_{f}}$, such that the state vector $x(t)$ of this descriptor system simultaneously satisfies $x(0)=x_{0}$ and $x(t_{f})=x_{f}$.

Regularity is a particular and important concept for a descriptor system. When a descriptor system is regular, uniqueness of its output is guaranteed, provided that it is stimulated by a consistent input.

The following results are well known about a descriptor system (\cite{Dai1989,Duan2010,it2017}).

\begin{lemma}
Assume that the descriptor system of Equation (\ref{eqn:6}) is regular. Then it is completely observable if and only if the following two conditions are satisfied simultaneously, %\vspace{-0.25cm}
\begin{itemize}
\item the matrix $\left[ E^{T} \;\; C^{T} \right]^{T}$ is of FCR;
\item the matrix pencil $\left[\lambda E^{T}\! -\! A^{T} \; C^{T} \right]^{T}$ is of FCR at each $\lambda\!\in\!{\cal C}$.
\end{itemize}
Moreover, it is completely controllable if and only if the following two conditions are satisfied simultaneously,  % \vspace{-0.25cm}
\begin{itemize}
\item the matrix $\left[ E \;\; B \right]$ is of FRR;
\item the matrix pencil $\left[\lambda E - A \;\; B \right]$ is of FRR at every $\lambda\in{\cal C}$.
\end{itemize}
\label{lemma:2}
\end{lemma}

Obviously, similar to a state space model, complete controllability and complete observability of a descriptor system are also dual to each other.

The following results are useful in this study about regularity, complete controllability/observability of the NDS $\rm\bf\Sigma$, which are well known in matrix analysis (\cite{Gantmacher1959,hj1991}).

\begin{lemma}
Partition a matrix $M$ as $M = \left[ M_{1}^{T} \;\;  M_{2}^{T}\right]^{T}$. Assume that the submatrix $M_{1}$ is not of FCR. Then the matrix $M$ is of FCR, if and only if the matrix $M_{2}M_{1}^{\perp}$ is of FCR.
\label{lemma:3}
\end{lemma}

On the basis of these conclusions, the following results are obtained in \cite{Zhou2019}, which are quite helpful in exploiting the block diagonal structure of the associated matrices in developing a computationally feasible verification procedure for the regularity/controllability/observability of the NDS $\rm\Sigma$.

\begin{lemma}
Let $A_{i}|_{i=1}^{3}$ and $B_{i}|_{i=1}^{3}$ be some matrices with compatible dimensions. Assume that the matrix $A_{2}$ is of FCR. Then the matrix
$\left[\begin{array}{c}
{\rm\bf diag}\left\{A_{1},\;A_{2},\; A_{3}\right\} \\ \left[ B_{1} \;\;\;\; B_{2} \;\;\;\; B_{3}\right]\end{array}\right]$ is of FCR, if and only if the
matrix
$\left[\begin{array}{c}
{\rm\bf diag}\left\{A_{1},\; A_{3}\right\} \\ \left[ B_{1} \;\; \;\; B_{3}\right]\end{array}\right]$ is.
\label{lemma:4}
\end{lemma}

The next results play an important role in revealing properties of the obtained conditions for NDS regularity and complete controllability/observability.

\begin{lemma}
Partition a matrix $M$ as $M = \left[M_{1}^{T} \;\;  M_{2}^{T}\right]^{T}$. Then
\begin{displaymath}
{\rm\bf null}(M)=M_{1}^{\perp}{\rm\bf null}(M_{2}M_{1}^{\perp})=M_{2}^{\perp}{\rm\bf null}(M_{1}M_{2}^{\perp})
\end{displaymath}
\label{lemma:5}
\end{lemma}

\hspace*{-0.4cm}{\it\bf Proof:} Assume that $\alpha\in {\rm\bf null}(M)$. Then $M_{1}\alpha=0$ and $M_{2}\alpha=0$. The first equation means that there exists a vector $\xi$, such that $\alpha=M_{1}^{\perp}\xi$. Substitute this expression into the second equation, we have that
\begin{equation}
M_{2}M_{1}^{\perp}\xi=0
\end{equation}
That is, $\xi \in {\rm\bf null}(M_{2}M_{1}^{\perp})$. Hence, $\alpha \in M_{1}^{\perp}{\rm\bf null}(M_{2}M_{1}^{\perp})$. This means that
${\rm\bf null}(M)\subseteq M_{1}^{\perp}{\rm\bf null}(M_{2}M_{1}^{\perp})$.

On the contrary, assume that $\alpha \in M_{1}^{\perp}{\rm\bf null}(M_{2}M_{1}^{\perp})$. Then there is a vector $\xi$ satisfying simultaneously $M_{2}M_{1}^{\perp}\xi=0$ and $\alpha=M_{1}^{\perp}\xi$. Therefore
\begin{equation}
M\alpha = \left[\begin{array}{c} M_{1}M_{1}^{\perp}\xi \\  M_{2}M_{1}^{\perp}\xi\end{array}\right]=0
\end{equation}
That is, $\alpha\in {\rm\bf null}(M)$, which further implies that $M_{1}^{\perp}{\rm\bf null}(M_{2}M_{1}^{\perp})\subseteq {\rm\bf null}(M)$.

It can therefore be declared that ${\rm\bf null}(M)=M_{1}^{\perp}{\rm\bf null}(M_{2}M_{1}^{\perp})$. The second equality of the lemma can be proved similarly. This completes the proof. \hspace{\fill}$\Diamond$

\section{Conrollability/Observability of the NDS}

For brevity, let $p$ denote the vector ${\rm\bf col}\left\{p_{i}|_{i=1}^{N}\right\}$.
Moreover, for ${\rm\bf \#}={\rm\bf x}$, ${\rm\bf v}$, or ${\rm\bf z}$, define a vector $\#(t)$ as $\#(t)={\rm\bf col}\left\{\#(t,i)|_{i=1}^{N}\right\}$. Furthermore, define matrices $D_{\rm\bf u}(p)$ and $E(p)$ respectively as $D_{\rm\bf u}(p)\!\!=\!\!{\rm\bf diag}\!\left\{D_{\rm\bf u}(i,p_{i})|_{i=1}^{N}\!\right\}$ and $E(p)\!\!=\!\!{\rm\bf diag}\!\left\{E(i,p_{i})|_{i=1}^{N}\!\right\}$.  In addition, define matrices $A_{\rm\bf
*\#}(p)$, $B_{\rm\bf *}(p)$ and $C_{\rm\bf *}(p)$ with ${\rm\bf *,\#}={\rm\bf x}$, ${\rm\bf y}$, ${\rm\bf v}$, or ${\rm\bf z}$, respectively, as $A_{\rm\bf
*\#}(p)=\!\!{\rm\bf diag}\!\left\{A_{\rm\bf
*\#}(i,p_{i})|_{i=1}^{N}\!\right\}$, $B_{\rm\bf *}(p)\!\!=\!\!{\rm\bf diag}\!\left\{B_{\rm\bf
*}(i,p_{i})|_{i=1}^{N}\!\right\}$ and $C_{\rm\bf *}(p)\!=\!{\rm\bf diag}\!\left\{C_{\rm\bf
*}(i,p_{i})|_{i=1}^{N}\!\right\}$.

Using these symbols, the dynamics of all the subsystems of the NDS $\rm\bf\Sigma$ can be compactly expressed as
\begin{equation}
\left[\! \begin{array}{c}
{E(p)\delta({x}(t))}\\
{{{z}}(t)}\\
{{y}(t)}
\end{array} \!\right] = \left[\! \begin{array}{ccc}
{A_{\rm\bf xx}{(p)}} & {A_{\rm\bf xv}{(p)}} & {B_{\rm\bf x}{(p)}}\\
{A_{\rm\bf zx}{(p)}} & {A_{\rm\bf zv}{(p)}} & {B_{\rm\bf z}{(p)}}\\
{C_{\rm\bf x}{(p)}} & {C_{\rm\bf v}{(p)}} & {D_{\rm\bf u}{(p)}}
\end{array} \!\right]\left[\! \begin{array}{c}
{x(t)}\\
{{v}(t)}\\
{u(t)}
\end{array} \!\right]
\label{eqn:1-c}
\end{equation}
Combining this equation with Equation (\ref{eqn:2}), a descriptor form can be obtained for the dynamics of the NDS $\rm\bf\Sigma$ that has completely the same form as that of Equation (\ref{eqn:6}), in which the matrices $A$, $B$, $C$, $D$ and $E$ are respectively replaced by the matrices $A(p)$, $B(p)$, $C(p)$, $D(p)$ and $E(p)$ with $A(p)$, $B(p)$, $C(p)$ and $D(p)$ being defined respectively as
\begin{eqnarray}
\hspace*{-0.5cm}\left[\!\!\begin{array}{cc} A(p) & B(p) \\  C(p) & D(p)
\end{array}\!\!\right] \!\!\!\!&=&\!\!\!\!\left[\!\!\begin{array}{cc}
A_{\rm\bf xx}(p) & \hspace*{-0.2cm} B_{\rm\bf x}(p) \\
C_{\rm\bf x}(p)  & \hspace*{-0.2cm} D_{\rm\bf u}(p) \end{array}\!\!\right]+\left[\!\!\begin{array}{c}
A_{\rm\bf xv}(p) \\
C_{\rm\bf v}(p)\end{array}\!\!\right]\times  \nonumber\\
& & \hspace*{-0.2cm} \Phi\!
\left[\;I\!-\!A_{\rm\bf zv}(p)\Phi\;\right]^{\!-1}\!\left[
A_{\rm\bf zx}(p)\;\; B_{\rm\bf z}(p)\right]
\label{eqn:7}
\end{eqnarray}

With these expressions, a necessary and sufficient condition is obtained for the regularity of the NDS $\rm\bf\Sigma$. Its proof is given in the appendix.

\begin{theorem}
Define a matrix pencil $\Theta(\lambda)$ as
\begin{equation}
{\Theta}(\lambda)=\left[\begin{array}{cc}
\lambda E(p) -A_{\rm\bf xx}(p) & -A_{\rm\bf xv}(p) \\
-{\Phi} A_{\rm\bf zx}(p) & I_{M_{\rm\bf v}}-{\Phi} A_{\rm\bf zv}(p) \end{array}\right]
\label{eqn:8}
\end{equation}
and let ${\rm\bf\Lambda}_{r}$ be a set of $M_{\rm\bf x}+1$ arbitrary but distinguished complex numbers. Then the NDS ${\rm\bf\Sigma}$ is regular if and only if there exists a $\lambda_{0}\in{\rm\bf \Lambda}_{r}$, such that the matrix $\Theta(\lambda_{0})$ is of FCR.
\label{theorem:1}
\end{theorem}

From the lumped descriptor form representation for the input-output relations of the NDS $\rm\bf\Sigma$, the following results have been established for its complete observability. The essential ideas behind the derivations are similar to those of \cite{Zhou2015,zyl2018}. That is, exploiting the LFT structure of the system matrices. The proof is deferred to the appendix.

\begin{theorem}
Define a matrix pencil ${\Xi}^{[o]}(\lambda)$ and a matrix ${\Xi}_{\infty}^{[o]}$ respectively as
\begin{eqnarray}
& &\hspace*{-0.5cm} {\Xi}^{[o]}(\lambda)=\left[\begin{array}{cc}
\lambda E(p)-A_{\rm\bf xx}(p) & -A_{\rm\bf xv}(p) \\
-C_{\rm\bf x}(p) & -C_{\rm\bf v}(p) \\
-\Phi A_{\rm\bf zx}(p) & I_{M_{\rm\bf v}}-\Phi A_{\rm\bf zv}(p) \end{array}\right]
\label{eqn:9}    \\
& & \hspace*{-0.5cm}{\Xi}_{\infty}^{[o]}=\left[\begin{array}{cc}
E(p) & 0 \\
-C_{\rm\bf x}(p) & -C_{\rm\bf v}(p) \\
-\Phi A_{\rm\bf zx}(p) & I_{M_{\rm\bf v}}-\Phi A_{\rm\bf zv}(p) \end{array}\right]
\label{eqn:10}
\end{eqnarray}
Assume that the NDS ${\rm\bf\Sigma}$ is regular. Then it is completely observable, if and only if the following two conditions are satisfied simultaneously,  % \vspace{-0.25cm}
\begin{itemize}
\item at every $\lambda\in {\cal C}$, the matrix pencil ${\Xi}^{[o]}(\lambda)$ is of FCR;
\item the matrix ${\Xi}_{\infty}^{[o]}$ is of FCR.
\end{itemize}
\label{theorem:2}
\end{theorem}

From Lemma \ref{lemma:2}, it is clear that complete controllability of a descriptor system is dual to its completely observability, which is well known in the analysis and synthesis of descriptor systems. On the basis of this relation, the following results are immediately obtained for the complete controllability of the NDS $\rm\bf\Sigma$. The proof is omitted due to its obviousness.

\begin{corollary}
Define a matrix pencil ${\Xi}^{[c]}(\lambda)$ and a matrix ${\Xi}_{\infty}^{[c]}$ respectively as
\begin{eqnarray}
& & \hspace*{-1.0cm}{\Xi}^{[c]}(\lambda) \!\!=\!\! \left[\!\!\!\begin{array}{ccc}
\lambda E(p)\!-\!A_{\rm\bf xx}(p) & -B_{\rm\bf x}(p) & -A_{\rm\bf xv}(p)\Phi \\
-A_{\rm\bf zx}(p) & -B_{\rm\bf z}(p) & I_{M_{\rm\bf z}}\!-\! A_{\rm\bf zv}(p)\Phi \end{array}\!\!\!\right]
\label{eqn:14}    \\
& & \hspace*{-1.0cm}{\Xi}_{\infty}^{[c]}=\left[\begin{array}{ccc}
 E(p)  & -B_{\rm\bf x}(p) & -A_{\rm\bf xv}(p)\Phi \\
0 & -B_{\rm\bf z}(p) & I_{M_{\rm\bf z}}- A_{\rm\bf zv}(p)\Phi \end{array}\right]
\label{eqn:15}
\end{eqnarray}
Assume that the NDS ${\rm\bf\Sigma}$ is regular. Then it is completely controllable,  if and only if the following two conditions are satisfied simultaneously,  % \vspace{-0.25cm}
\begin{itemize}
\item at every $\lambda\in{\cal C}$, the matrix pencil $\Xi^{[c]}(\lambda)$ is of FRR;
\item the matrix $\Xi_{\infty}^{[c]}$ is of FRR.
\end{itemize}
\label{corollary:1}
\end{corollary}

Theorem \ref{theorem:1} gives a necessary and sufficient condition for the regularity of the NDS $\rm\bf\Sigma$, while Theorem \ref{theorem:2} and Corollary \ref{corollary:1} respectively a necessary and sufficient condition for its complete observability and complete controllability. However, these conditions are still not computationally feasible, noting that in these conditions, the rank must be checked for the matrix pencil ${\Xi}^{[o]}(\lambda)$/${\Xi}^{[c]}(\lambda)$ at infinitely many values of the complex variable $\lambda$ that is computationally prohibitive. On the other hand, when a large scale NDS is under investigation, the dimension of the matrix $\Theta(\lambda)$ is generally high at any $\lambda\in{\cal C}$, which is also not computationally attractive.

When the parameters of each subsystem are known, all the matrices involved in Theorems \ref{theorem:1} and \ref{theorem:2}, as well as Corollary \ref{corollary:1}, that is, the matrices $A_{\rm\bf xx}(p)$, $A_{\rm\bf xv}(p)$, etc., are also known. To develop a computationally feasible condition from the above results for the complete observability of the NDS $\rm\Sigma$ under this situation, assume that for each $i=1,2,\cdots,N$, the matrix $\left[C_{\rm\bf x}(i)\;\; C_{\rm\bf v}(i)\right]$ is column rank deficient. This assumption is adopted only for notational simplicity and does not affect validity of the results obtained in this paper. In particular, if there is a subsystem that does not satisfy this assumption, then it is clear from Lemma \ref{lemma:4} that, removing all system matrices of this subsystem from the matrix pencil
${\Xi}^{[o]}(\lambda)$ of Equation (\ref{eqn:9}) and the matrix ${\Xi}^{[o]}_{\infty}$ of Equation (\ref{eqn:10}), does not change conclusions on NDS complete observability. Under this assumption, the right null space of the matrix $\left[C_{\rm\bf x}(i)\;\; C_{\rm\bf v}(i)\right]$ is not a zero vector. Moreover, there exist matrices $N_{\rm\bf x}(i)$ and $N_{\rm\bf v}(i)$ satisfying
\begin{equation}
{\rm\bf Null}\left(\left[C_{\rm\bf x}(i)\;\; C_{\rm\bf v}(i)\right]\right)
={\rm\bf Span}\left(\left[\begin{array}{c} N_{\rm\bf x}(i)\\ N_{\rm\bf v}(i)\end{array}\right]\right)
\label{eqn:16}
\end{equation}
in which the matrix ${\rm\bf col}\left\{ N_{\rm\bf x}(i),\;\; N_{\rm\bf v}(i)\right\}$ is of FCR. In addition, the matrices $N_{\rm\bf x}(i)$ and $N_{\rm\bf v}(i)$ have a dimension compatible with those of the matrices $C_{\rm\bf x}(i)$ and $C_{\rm\bf v}(i)$.

In order to simplify expressions, the dependence of the matrix $C_{\rm\bf x}(i,p_{i})$, etc. on the parameter vector $p_{i}$ is omitted in the above paragraph. This omission is adopted in the rest of this section. On the other hand, if there is a subsystem with the associated $E(i,p_{i})$ not being of FRR, some elementary manipulations on its rows can transfer the NDS $\rm\bf\Sigma$ into another one without changing conclusions about its complete observability, in which each of the matrices $E(i,p_{i})|_{i=1}^{N}$ is of FRR. These apply also to the following matrix $E(i)N_{\rm\bf x}(i)$.

From Lemma \ref{lemma:1}, it can be declared that for each $i\in\{1,2,\cdots,N\}$, there exist two invertible constant real matrices $U(i)$ and $V(i)$, some unique nonnegative integers $\xi_{Hi}$, $\zeta_{Ki}$, $\zeta_{Ni}$, $\zeta_{Li}$, $\zeta_{Ji}$, $\xi_{Li}(j)|_{j=1}^{\zeta_{Li}}$ and $\xi_{Ji}(j)|_{j=1}^{\zeta_{Ji}}$, some unique positive integers $\xi_{Ki}(j)|_{j=1}^{\zeta_{Ki}}$ and $\xi_{Ni}(j)|_{j=1}^{\zeta_{Ni}}$, as well as a strictly regular $\xi_{Hi}\times \xi_{Hi}$ dimensional matrix pencil $H_{\xi_{Hi}}(\lambda)$, such that
\begin{eqnarray}
& & \!\!\lambda E(i)N_{\rm\bf x}(i)-\left[A_{\rm\bf xx}(i)N_{\rm\bf x}(i)+A_{\rm\bf xv}(i)N_{\rm\bf v}(i)\right] \nonumber\\
&=&\!\!U(i){\rm\bf diag}\!\left\{\!H_{\xi_{Hi}}(\lambda),\;K_{\xi_{Ki}(j)}(\lambda)|_{j=1}^{\zeta_{Ki}},\; L_{\xi_{Li}(j)}(\lambda)|_{j=1}^{\zeta_{Li}}, \right.\nonumber \\
& & \hspace*{1.0cm}\left. N_{\xi_{Ni}(j)}(\lambda)|_{j=1}^{\zeta_{Ni}},\; J_{\xi_{Ji}(j)}(\lambda)|_{j=1}^{\zeta_{Ji}}\!\right\}V(i)
\label{eqn:17}
\end{eqnarray}

Let $n(i)$ and $m(i)$ denote respectively $\xi_{Hi}+\sum_{j=1}^{\zeta_{Ki}}\xi_{Ki}(j)+1$ and $\zeta_{Li}+\xi_{Hi}+\sum_{j=1}^{\zeta_{Ki}}\xi_{Ki}(j)+\sum_{j=1}^{\zeta_{Li}}\xi_{Li}(j)$ for each $i=1,2,\cdots,N$. Moreover, let $V_{i}^{-1}(m(i))$ and $V_{i}^{-1}(n(i):m(i))$ respectively represent the matrix constructed by the first $m(i)$ columns of the inverse of the matrix $V(i)$ and that from its $n(i)$-th column to its $m(i)$-th column. On the other hand, let $\rm\bf\Lambda$ stand for the set of complex numbers at which the matrix pencil % \vspace{-0.25cm}
\begin{equation} %\vspace{-0.25cm}
{\rm\bf diag}\!\left\{\left.{\rm\bf diag}\!\left\{\!H_{\xi_{Hi}}(\lambda),\;K_{\xi_{Ki}(j)}(\lambda)|_{j=1}^{\zeta_{Ki}}\right\}\right|_{i=1}^{N}\right\}
\label{eqn:18}
\end{equation}
is not of FCR. Recall that a matrix pencil in the form $H_{\star}(\lambda)$ or $K_{\star}(\lambda)$ is column rank deficient only at finitely many isolated values of its variable $\lambda$. It is obvious that the set $\rm\bf\Lambda$ only has finitely many elements. In addition, for each $i\in\{1,2,\cdots,N\}$ and each $\lambda_{0}\in{\rm\bf\Lambda}$, let $N(\lambda_{0},i)$ denote a matrix with independent columns that span the null space of the following matrix % \vspace{-0.25cm}
\begin{displaymath}
% \vspace{-0.25cm}
{\rm\bf diag}\!\left\{\!H_{\xi_{Hi}}(\lambda_{0}),\;K_{\xi_{Ki}(j)}(\lambda_{0})|_{j=1}^{\zeta_{Ki}},\; L_{\xi_{Li}(j)}(\lambda_{0})|_{j=1}^{\zeta_{Li}}\right\}
\end{displaymath}
and denote $V_{i}^{-1}(m(i))N(\lambda_{0},i)$ by $\bar{N}(\lambda_{0},i)$. Furthermore, define an MVP $\Upsilon(\lambda,i)$ as %\vspace{-0.25cm}
\begin{eqnarray*} %\vspace{-0.25cm}
\Upsilon(\lambda,i)&=& V_{i}^{-1}(n(i):m(i))\times\\
& &
{\rm\bf diag}\!\left\{ \left.{\rm\bf col}\left\{1,\;\lambda,\; \cdots,\; \lambda^{\xi_{Li}(j)}\right\}\right|_{j=1}^{\zeta_{Li}}\right\}
\end{eqnarray*}

Using these symbols, the following results are obtained from Theorem \ref{theorem:2}, as well as Lemmas \ref{lemma:3} and \ref{lemma:4}. Their proof is deferred to the appendix.

\begin{theorem}
For each $\lambda_{0}\in{\rm\bf\Lambda}$, define matrices $X(\lambda_{0})$ and $Y(\lambda_{0})$ respectively as
\begin{eqnarray*}
& & \hspace*{-0.5cm} X(\lambda_{0})\!=\!{\rm\bf diag}\!\left\{\!\left.
N_{\rm\bf v}(i)\bar{N}(\lambda_{0},i)
\right|_{i=1}^{N}\right\}  \\
& & \hspace*{-0.5cm} Y(\lambda_{0})\!=\!{\rm\bf diag}\!\left\{\!\left.\left[A_{\rm\bf zx}(i)N_{\rm\bf x}(i)\!+\!A_{\rm\bf zv}(i)N_{\rm\bf v}(i)\right]\!\bar{N}(\lambda_{0},i)
\right|_{i=1}^{N}\!\right\}
\end{eqnarray*}
Moreover, define MVPs $\Omega(\lambda)$ and $\Gamma(\lambda)$ respectively as
\begin{eqnarray*}
& & \hspace*{-0.5cm} \Omega(\lambda)\!=\!{\rm\bf diag}\!\left\{\!\left.
N_{\rm\bf v}(i)\Upsilon(\lambda,i)
\right|_{i=1}^{N}\right\}  \\
& & \hspace*{-0.5cm} \Gamma(\lambda)\!=\!{\rm\bf diag}\!\left\{\!\left.\left[A_{\rm\bf zx}(i)N_{\rm\bf x}(i)\!+\!A_{\rm\bf zv}(i)N_{\rm\bf v}(i)\right]\!\Upsilon(\lambda,i)
\right|_{i=1}^{N}\!\right\}
\end{eqnarray*}
Then the matrix pencil ${\Xi}^{[o]}(\lambda)$ of Theorem \ref{theorem:2} is of FCR at each $\lambda\in {\cal C}$, if and only if the following two conditions are satisfied simultaneously, % \vspace{-0.25cm}
\begin{itemize}
\item the matrix $X(\lambda_{0})-\Phi Y(\lambda_{0})$ is of FCR at each $\lambda_{0}\in{\rm\bf\Lambda}$;
\item the MVP $\Omega(\lambda)-\Phi\Gamma(\lambda)$ is always of FCR.
\end{itemize}
\label{theorem:4}
\end{theorem}

Obviously, if there exists an integer $i$ belonging to the set $\{1,2,\cdots,N\}$, such that $\zeta_{Li}\neq 0$, then the 2nd condition of the above theorem becomes active. This means that under such a situation, in order to construct a completely observable NDS $\rm\bf\Sigma$, its SCM $\Phi$ must satisfy infinitely many constraints. This is clearly not attractive in actual applications.

When the NDS SCM $\Phi$ is known, the 2nd condition of Theorem \ref{theorem:4} can in principle be verified using the Smith form of a MVP. Note that the MVP $\Omega(\lambda)-\Phi\Gamma(\lambda)$ has $\sum_{i=1}^{N}\zeta_{Li}$ columns and its degree is equal to $\max_{1\leq i\leq N}\max_{1\leq j\leq \zeta_{Li}} \xi_{Li}(j)$. The computations are in general quite intensive when the number of $L_{\star}(\lambda)$ form matrix pencils in Equation (\ref{eqn:17}) are large and/or some of them have a high dimension.

From their definitions, it can be seen that for any particular $\lambda_{0}\in{\cal C}$, the associated matrices $X(\lambda_{0})$ and $Y(\lambda_{0})$ can be calculated from each subsystem individually. This property also holds for the set $\rm\bf\Lambda$ and the MVPs $\Omega(\lambda)$ and $\Gamma(\lambda)$, which makes the associated conditions attractive in the analysis and synthesis of a large scale NDS. On the other hand, as the set $\rm\bf\Lambda$ is determined independently by each subsystem, it appears possible to utilize this property in subsystem designs such that a completely observable NDS can be constructed more easily.

In case that the subsystem dynamics of the NDS $\rm\bf\Sigma$ are homogeneous, which has been extensively investigated under a name like multi-agent system, Roesser model and FM model, etc. (\cite{cam2014,fm1978,np2013,Roesser1975,sbkkmpr2011,Zhou2006,zyl2018}), applications of the Kronecker product can further reduce computational costs and reveal relations between NDS complete observability and its subsystem parameters/connections. More specifically, for each $i\in\{1,2,\cdots,N\}$ and each $\lambda_{0}\in{\rm\bf\Lambda}$, denote
$N_{\rm\bf v}(i)\bar{N}(\lambda_{0},i)$ and $\left[A_{\rm\bf zx}(i)N_{\rm\bf x}(i)\!+\!A_{\rm\bf zv}(i)N_{\rm\bf v}(i)\right]\!\bar{N}(\lambda_{0},i)$ respectively by $X(\lambda_{0},i)$ and $Y(\lambda_{0},i)$. Then it is not difficult to see that under the aforementioned homogeneity assumption, $X(\lambda_{0},1)=X(\lambda_{0},2)=\cdots=X(\lambda_{0},N)$ and $Y(\lambda_{0},1)=Y(\lambda_{0},2)=\cdots=Y(\lambda_{0},N)$, which further leads to $X(\lambda_{0})=I_{N}\otimes X(\lambda_{0},1)$ and
$Y(\lambda_{0})=I_{N}\otimes Y(\lambda_{0},1)$, in which $\otimes$ stands for the Kronecker product. Assume that there is a real and constant matrix $\Phi(1)$ such that the SCM $\Phi$ satisfies $\Phi=I_{N}\otimes \Phi(1)$, which means that each subsystem is influenced by its neighbors through the same way that is satisfied by the Roesser model, the FM model, etc. Then using properties of Kronecker products (\cite{Gantmacher1959,hj1991}), it can be straightforwardly proven that
\begin{displaymath}
X(\lambda_{0})-\Phi Y(\lambda_{0})=I_{N}\otimes \left[X(\lambda_{0},1)-\Phi(1)Y(\lambda_{0},1)\right]
\end{displaymath}
Hence, $X(\lambda_{0})-\Phi Y(\lambda_{0})$ is of FCR if and only if $X(\lambda_{0},1)-\Phi(1) Y(\lambda_{0},1)$ is of FCR. Similar conclusions can be achieved for the 2nd condition of Theorem \ref{theorem:4}. Hence, NDS complete observability is determined completely by its individual subsystem dynamics and connections, regardless of how many subsystems it has. These declarations apply also to the conclusions about NDS complete controllability and those about subsystem parameter influences given in the next section.

Through similar arguments, conditions can be derived respectively for the verification of the nonsingularity of the matrix pencil $\Theta(\lambda)$ of Equation (\ref{eqn:8}) at any prescribed $\lambda_{0}\in{\cal C}$, and for verifying whether or not the matrix ${\Xi}_{\infty}^{[o]}$ defined by Equation (\ref{eqn:10}) is of FCR. These conditions have completely the same form, as well as the same properties, as that of Theorem \ref{theorem:4}. More precisely, the null space of the matrix
$\left[\lambda_{0} E(p) -A_{\rm\bf xx}(p) \;\;  -A_{\rm\bf xv}(p) \right]$ can be constructed from each subsystem individually with a given $\lambda_{0}\in {\cal C}$ and given values of the subsystem parameters $p_{i}|_{i=1}^{N}$. With this null space and Lemma \ref{lemma:3}, a necessary and sufficient condition can be obtained for the invertibility of the matrix $\Theta(\lambda_{0})$, which is similar to the 1st condition of the aforementioned theorem. On the other hand, on the basis of Lemma \ref{lemma:5}, the null space of the matrix $\left[\begin{array}{cc} E(p) & 0 \\ -C_{\rm\bf x}(p) & -C_{\rm\bf v}(p)  \end{array}\right]$ can also be constructed from each subsystem independently. By means of this null space and Lemma \ref{lemma:3}, a necessary and sufficient condition can be obtained for the matrix ${\Xi}_{\infty}^{[o]}$ being of FCR, which is again similar to the 1st condition of the aforementioned theorem. The details are omitted due to space considerations.

By means of the same token, computationally feasible conditions can also be obtained for the complete controllability of the NDS $\rm\bf\Sigma$, provided that the parameter vector is known for each of its subsystems. The associated conclusions and derivations are not included for their obviousness and space considerations.

\section{Dependence of System Observability on Subsystem Parameters}

The previous section has made it clear that in order to construct a completely controllable/observable NDS, characteristics of its subsystems are quite important. Particularly, if a subsystem is not selected appropriately, it will lead to an infinite number of constraints on the SCM of the NDS. However, it is still impossible to investigate the influence of a subsystem parameter on the regularity/controllability/observability of the whole NDS under the general situation, that is, when a subsystem parameter affects subsystem matrices through an arbitrary function. As a first step, it is assumed throughout this section that
\begin{eqnarray}
& & \hspace*{-0.8cm}\left[\!\! \begin{array}{c}
E(i,p_{i}) \\
{A_{\rm\bf xx}{(i,p_{i})}} \\
{A_{\rm\bf zx}{(i,p_{i})}} \\
{C_{\rm\bf x}{(i,p_{i})}} \end{array}\!\!\right]\!\!=\!\!
 \left[\!\! \begin{array}{c}
E^{[0]}(i) \\
{A_{\rm\bf xx}^{[0]}{(i)}} \\
A_{\rm\bf zx}^{[0]}{(i)} \\
C_{\rm\bf x}^{[0]}{(i)} \end{array}\!\!\right] \!+\!
\left[\! \begin{array}{c}
F_{1}(i)\\ F_{2}(i) \\ F_{3}(i)  \\ F_{4}(i) \end{array}\!\! \right] \!
\left[ M(i)\!-\!P_{1}(i)H(i)\right]^{-1}\!\!\times \nonumber\\
& & \hspace*{5.0cm} P_{1}(i)G(i)
\label{eqn:1-a}  \\
& & \hspace*{-0.8cm} \left[\!\! \begin{array}{c}
{A_{\rm\bf xv}{(i,p_{i})}} \\
{A_{\rm\bf zv}{(i,p_{i})}} \\
{C_{\rm\bf v}{(i,p_{i})}} \end{array}\!\!\right]\!\!=\!\!
 \left[\!\! \begin{array}{c}
{A_{\rm\bf xv}^{[0]}{(i)}} \\
{A_{\rm\bf zv}^{[0]}{(i)}} \\
{C_{\rm\bf v}^{[0]}{(i)}} \end{array}\!\!\right] \!+\!
\left[\!\!\begin{array}{c}
J_{1}(i)\\ J_{2}(i) \\ J_{3}(i)  \end{array}\!\! \right] \!
\left[ N(i)\!-\!P_{2}(i)S(i)\right]^{-1}\!\!\times \nonumber\\
& & \hspace*{5.0cm} P_{2}(i)K(i)
\label{eqn:1-b}
\end{eqnarray}

In the above equations, the matrices $P_{1}(i)$ and $P_{2}(i)$ consist of elements that are constantly equal to zero and elements that can be expressed as a function of the elements of the parameter vector $p_{i}$. The matrices $G{(i)}$, $H{(i)}$, $M{(i)}$, $K{(i)}$, $S(i)$ and $N{(i)}$, together with the matrices $F_j{(i)}$ and the matrices $J_j{(i)}$ with $j=1,2,3,4$, are matrices reflecting how this subsystem's FPPs affect its system matrices. These matrices, together with the matrices ${E^{[0]}(i)}$, ${A_{\rm\bf *\#}^{[0]}(i)}$ and ${C_{\rm\bf *}^{[0]}(i)}$ with ${\rm\bf *,\#}={\rm\bf x}$, ${\rm\bf v}$ or ${\rm\bf z}$, are in general known and can not be selected or adjusted in system designs, as they reflect the physical, chemical, electrical or other principles governing the dynamics of this subsystem, such as the Kirchhoff's current law, Newton's mechanics, etc.

The expressions of Equations (\ref{eqn:1-a}) and (\ref{eqn:1-b}) are essentially in the
form of $\Psi_{22}+\Psi_{21}\left[\Xi-\Delta\Psi_{11}\right]^{-1}\Delta\Psi_{12}$. A transformation in this form with a fixed matrix $\Xi$ and some fixed matrices $\Psi_{ij}|_{i,j=1}^{2}$ is called a GLFT of the matrix $\Delta$, which is originally introduced in \cite{hv2004} for parametric uncertainty descriptions. It is argued there that although both an LFT and a GLFT are capable of expressing an arbitrary rational function, a GLFT usually has a lower matrix dimension than an LFT in the associated expressions. As computational cost is one of the most essential issues in the analysis and synthesis of a large scale NDS, rather than LFTs which are adopted in \cite{Zhou2019}, it is the above GLFT that is adopted in this paper in describing the dependence of system matrices of a subsystem in the NDS $\rm\bf\Sigma$ on its parameters.

In the above description, the matrices $P_{1}(i)$ and $P_{2}(i)$ consist of fixed zero elements and elements which are from the set consisting of all the FPPs of the subsystem ${\bf{\Sigma}}_i$, $i=1,2,\cdots,N$. In some situations, it may be more convenient to use a simple function of some FPPs, such as the reciprocal of a FPP, the product of several FPPs, etc. These transformations do not affect results of this paper, provided that the corresponding global transformation is an injective mapping. To avoid an awkward presentation, these elements are called pseudo FPP in this paper, and are usually assumed to be algebraically independent of each other.

Note that in the adopted model, the matrix $E(i,p_{i})$ also depends on the subsystem parameter vector $p_{i}$. This situation happens in actual applications (\cite{hv2004}), and disables the approach adopted in \cite{Zhou2019}, in which the inputs and outputs of each subsystem, as well as the SCM of the NDS, are augmented such that the parameters of each subsystem are included in an augmented SCM, and the resulted NDS takes the same form as an NDS without any parameters or with each parameter being prescribed.

Define matrices $P_{1}$, $P_{2}$, $G$, $K$, $M$ and $N$ respectively as $P_{1}\!\!=\!\!{\rm\bf diag}\!\left\{\!P_{1}(i)|_{i=1}^{N}\!\right\}$, $P_{2}\!\!=\!\!{\rm\bf diag}\!\left\{\!P_{2}(i)|_{i=1}^{N}\!\right\}$, $G\!\!=\!\!{\rm\bf diag}\!\left\{\!G(i)|_{i=1}^{N}\!\right\}$, $K\!\!=\!\!{\rm\bf diag}\!\left\{\!K(i)|_{i=1}^{N}\!\right\}$, $M\!\!=\!\!{\rm\bf diag}\!\left\{\!M(i)|_{i=1}^{N}\!\right\}$ and $N\!\!=\!\!{\rm\bf diag}\!\left\{\!N(i)|_{i=1}^{N}\!\right\}$. Moreover, define matrices $F_{j}$ with $j=1,2,3,4$ and matrices $J_{j}$ with $j=1,2,3$ as
$F_{j}\!\!=\!\!{\rm\bf diag}\!\left\{\!F_{j}(i)|_{i=1}^{N}\!\right\}$
and $J_{j}\!\!=\!\!{\rm\bf diag}\!\left\{\!J_{j}(i)|_{i=1}^{N}\!\right\}$.
In addition, define matrices $A^{[0]}_{\rm\bf
*\#}\!\!$ and $C^{[0]}_{\rm\bf *}$ with ${\rm\bf *,\#}={\rm\bf x}$, ${\rm\bf v}$ or ${\rm\bf z}$, respectively as
$A^{[0]}_{\rm\bf
*\#}\!\!=\!\!{\rm\bf diag}\!\left\{\!A^{[0]}_{\rm\bf
*\#}(i)|_{i=1}^{N}\!\right\}$ and $C^{[0]}_{\rm\bf *}\!\!=\!\!{\rm\bf diag}\!\left\{\!C_{\rm\bf
*}(i)|_{i=1}^{N}\!\right\}$. If for each $i=1,2,\cdots,N$, the system matrices of the subsystem ${\rm\bf \Sigma}_{i}$ depend on its parameters through a way expressed by Equations (\ref{eqn:1-a}) and (\ref{eqn:1-b}), then the following results can be obtained, while their proof is given in the appendix.

\begin{theorem}
Define a matrix pencil $\Xi^{[o]}_{p}(\lambda)$ as
\begin{equation}
\Xi^{[o]}_{p}(\lambda)\!\!=\!\!\left[\!\!\!\!\begin{array}{cccc}
\lambda E^{[0]}\!-\!A_{\rm\bf xx}^{[0]} & \lambda F_{1}\!-\!F_{2} & -A_{\rm\bf xv}^{[0]} & -J_{1} \\
-C_{\rm\bf x}^{[0]} & -F_{3} & -C_{\rm\bf v}^{[0]}  & -J_{2} \\
-P_{1}G & M\!-\!P_{1}H & 0 & 0 \\
0 & 0 & -P_{2}K & N\!-\!P_{2}S \\
-\Phi A_{\rm\bf zx}^{[0]} & -\Phi F_{4} & I_{M_{\rm\bf v}}\!-\!\Phi A_{\rm\bf zv}^{[0]} & -\Phi J_{3} \end{array}\!\!\!\!\right]
\label{eqn:11}
\end{equation}
Then at any $\lambda\in{\cal C}$,
the matrix pencil $\Xi^{[o]}(\lambda)$ of Equation (\ref{eqn:9}) is of FCR, if and only if the matrix pencil $\Xi^{[o]}_{p}(\lambda)$ holds this property.
\label{theorem:3}
\end{theorem}

Note that the matrix pencil $\Xi^{[o]}_{p}(\lambda)$ of Equation (\ref{eqn:11}) has a similar structure as the matrix pencil $\Xi^{[o]}(\lambda)$ of Equation (\ref{eqn:9}) in which the parameter vector $p$ is prescribed. Through the adoption of the null space of the matrix $\left[C_{\rm\bf x}^{[0]} \;\; F_{3} \;\; C_{\rm\bf v}^{[0]} \;\; J_{2}\right]$ and utilization of Lemmas \ref{lemma:1} and \ref{lemma:3}, a rank condition similar to that of Theorem \ref{theorem:4} can be established for verifying whether or not the matrix pencil $\Xi^{[o]}_{p}(\lambda)$ is of FCR at each $\lambda\in{\cal C}$. In this condition, the associated matrix depends once again affinely on both the SCM $\Phi$ and the system parameter matrices $P_{1}$ and $P_{2}$.

More precisely, for each $i=1,2,\cdots,N$, assume that
\begin{eqnarray}
& &\!\!\!\! {\rm\bf Null}\!\left(\left[C^{[0]}_{\rm\bf x}(i)\;\; F_{3}(i)\;\; C^{[0]}_{\rm\bf v}(i)\;\; J_{2}(i)\right]\right) \nonumber\\
&=&\!\!\!\! {\rm\bf Span}\!\left(\!{\rm\bf col}\left\{ N^{[0]}_{\rm\bf x}(i),\;\;
N_{\rm\bf f}(i), \;\; N^{[0]}_{\rm\bf v}(i),\;\; N_{\rm\bf j}(i)\right\}\!\right)
\label{eqn:19}
\end{eqnarray}
Here, the matrices $N^{[0]}_{\rm\bf x}(i)$, $N_{\rm\bf f}(i)$, $N^{[0]}_{\rm\bf v}(i)$ and $N_{\rm\bf j}(i)$ are selected such that their dimensions are compatible respectively with those of the matrices $C^{[0]}_{\rm\bf x}(i)$, $F_{3}(i)$, $C^{[0]}_{\rm\bf v}(i)$ and $J_{2}(i)$, while the matrix ${\rm\bf col}\left\{ N^{[0]}_{\rm\bf x}(i),\;\; N_{\rm\bf f}(i),\;\;N^{[0]}_{\rm\bf v}(i),\;\;N_{\rm\bf j}(i)\right\}$ is of FCR. Existence of these matrices is guaranteed by matrix theories, provided that the matrix $\left[C^{[0]}_{\rm\bf x}(i)\;\; F_{3}(i)\;\; C^{[0]}_{\rm\bf v}(i)\;\; J_{2}(i)\right]$ is column rank deficient (\cite{Gantmacher1959,hj1991}). When this condition is not satisfied by a subsystem, approaches of the previous section must be adopted. That is, the  system matrices of the associated subsystem should be removed from the matrix pencil $\Xi^{[o]}_{p}(\lambda)$.

From Lemma \ref{lemma:1}, for each $i\in\{1,2,\cdots,N\}$, there exist two invertible constant real matrices $U^{[0]}(i)$ and $V^{[0]}(i)$, some unique nonnegative integers $\xi^{[0]}_{Hi}$, $\zeta^{[0]}_{Ki}$, $\zeta^{[0]}_{Ni}$, $\zeta^{[0]}_{Li}$, $\zeta^{[0]}_{Ji}$, $\xi^{[0]}_{Li}(j)|_{j=1}^{\zeta^{[0]}_{Li}}$ and
$\xi^{[0]}_{Ji}(j)|_{j=1}^{\zeta^{[0]}_{Ji}}$, some unique positive integers $\xi^{[0]}_{Ki}(j)|_{j=1}^{\zeta^{[0]}_{Ki}}$ and $\xi^{[0]}_{Ni}(j)|_{j=1}^{\zeta^{[0]}_{Ni}}$, as well as a strictly regular $\xi^{[0]}_{Hi}\times \xi^{[0]}_{Hi}$ dimensional matrix pencil $H_{\xi^{[0]}_{Hi}}(\lambda)$, such that
\begin{eqnarray}
& & \!\!\lambda \left[ E^{[0]}(i)N^{[0]}_{\rm\bf x}(i)+F_{1}(i)N_{\rm\bf f}(i)\right] - \left[ A^{[0]}_{\rm\bf xx}(i)N^{[0]}_{\rm\bf x}(i)+ \right. \nonumber\\
& & \hspace*{1.5cm}
\left. A^{[0]}_{\rm\bf xv}(i)N^{[0]}_{\rm\bf v}(i)+ F_{2}(i)N_{\rm\bf f}(i)+J_{1}(i)N_{\rm\bf j}(i)\right] \nonumber\\
&=&\!\!U^{[0]}(i){\rm\bf diag}\!\left\{\!H_{\xi^{[0]}_{Hi}}(\lambda),\;K_{\xi^{[0]}_{Ki}(j)}(\lambda)|_{j=1}^{\zeta^{[0]}_{Ki}},
L_{\xi^{[0]}_{Li}(j)}(\lambda)|_{j=1}^{\zeta^{[0]}_{Li}},\;\right.\nonumber \\
& & \hspace*{1.2cm}\left. N_{\xi^{[0]}_{Ni}(j)}(\lambda)|_{j=1}^{\zeta^{[0]}_{Ni}},\; J_{\xi^{[0]}_{Ji}(j)}(\lambda)|_{j=1}^{\zeta^{[0]}_{Ji}}\!\right\}\!\!V^{[0]}(i)
\label{eqn:20}
\end{eqnarray}

For each $i=1,2,\cdots,N$, let $m^{[0]}(i)$ denote $\zeta^{[0]}_{Li}+\xi^{[0]}_{Hi}+\sum_{j=1}^{\zeta^{[0]}_{Ki}}\xi^{[0]}_{Ki}(j)+\sum_{j=1}^{\zeta^{[0]}_{Li}}\xi^{[0]}_{Li}(j)$. Moreover, let $V_{i,0}^{-1}(m^{[0]}(i))$ represent the matrix consisting of the first $m^{[0]}(i)$ columns of the inverse of the matrix $V^{[0]}(i)$. Furthermore, let ${\rm\bf\Lambda}^{[0]}(i)$ stand for the set of complex numbers at which the following matrix pencil is not of FCR,
\begin{equation}
{\rm\bf diag}\!\left\{\!H_{\xi^{[0]}_{Hi}}(\lambda),\;K_{\xi^{[0]}_{Ki}(j)}(\lambda)|_{j=1}^{\zeta^{[0]}_{Ki}},\; L_{\xi^{[0]}_{Li}(j)}(\lambda)|_{j=1}^{\zeta^{[0]}_{Li}}\right\}
\label{eqn:21}
\end{equation}
Denote the set $\bigcup_{i=1}^{N}{\rm\bf\Lambda}^{[0]}(i)$ by ${\rm\bf\Lambda}^{[0]}$. Moreover, for each $1\leq i\leq N$ and each $\lambda_{0}\in{\rm\bf\Lambda}^{[0]}$, let $N^{[0]}(\lambda_{0},i)$ denote a matrix of FCR that spans the null space of the matrix
${\rm\bf diag}\!\left\{\!H_{\xi^{[0]}_{Hi}}(\lambda_{0}),\;K_{\xi^{[0]}_{Ki}(j)}(\lambda_{0})|_{j=1}^{\zeta^{[0]}_{Ki}},\; L_{\xi^{[0]}_{Li}(j)}(\lambda_{0})|_{j=1}^{\zeta^{[0]}_{Li}}\right\}$. Using these symbols, through similar arguments as those in the proof of Theorem \ref{theorem:4}, the following results can be established from Theorem \ref{theorem:3} and Lemmas \ref{lemma:3} and \ref{lemma:4}. Their proof is omitted due to obviousness.

\begin{theorem}
For each $\lambda_{0}\in{\rm\bf\Lambda}^{[0]}$ and each $i\in\{1,2,\cdots,N\}$, denote the matrix $V_{i,0}^{-1}(m^{[0]}(i))N^{[0]}(\lambda_{0},i)$ by $\bar{N}^{[0]}(\lambda_{0},i)$. Moreover, define matrices $X_{j}^{[0]}(\lambda_{0})$ and $Y_{j}^{[0]}(\lambda_{0})$ with $j=1,2,3$, respectively as
\begin{eqnarray*}
& & \hspace*{-0.5cm} X_{1}^{[0]}(\lambda_{0})\!=\!{\rm\bf diag}\!\left\{\!\left.
M(i)N_{\rm\bf f}(i)\bar{N}^{[0]}(\lambda_{0},i)
\right|_{i=1}^{N}\right\}  \\
& & \hspace*{-0.5cm} Y_{1}^{[0]}(\lambda_{0})\!=\!{\rm\bf diag}\!\left\{\!\left.\left[G(i)N^{[0]}_{\rm\bf x}(i)\!+\!H(i)N_{\rm\bf f}(i)\right] \bar{N}^{[0]}(\lambda_{0},i)
\right|_{i=1}^{N}\!\right\}  \\
& & \hspace*{-0.5cm} X_{2}^{[0]}(\lambda_{0})\!=\!{\rm\bf diag}\!\left\{\!\left.
N(i)N_{\rm\bf j}(i)\bar{N}^{[0]}(\lambda_{0},i)
\right|_{i=1}^{N}\right\}  \\
& & \hspace*{-0.5cm} Y_{2}^{[0]}(\lambda_{0})\!=\!{\rm\bf diag}\!\left\{\!\left.\left[K(i)N^{[0]}_{\rm\bf v}(i)\!+\!S(i)N_{\rm\bf j}(i)\right] \bar{N}^{[0]}(\lambda_{0},i)
\right|_{i=1}^{N}\!\right\}  \\
& & \hspace*{-0.5cm} X_{3}^{[0]}(\lambda_{0})\!=\!{\rm\bf diag}\!\left\{\!\left.
N^{[0]}_{\rm\bf v}(i)\bar{N}^{[0]}(\lambda_{0},i)
\right|_{i=1}^{N}\right\}  \\
& & \hspace*{-0.5cm} Y_{3}^{[0]}(\lambda_{0})\!=\!{\rm\bf diag}\!\left\{\!\left.\left[A^{[0]}_{\rm\bf zx}(i)N^{[0]}_{\rm\bf x}(i)\!+\!A^{[0]}_{\rm\bf zv}(i)N^{[0]}_{\rm\bf v}(i)\!+\!F_{4}(i)\times \right.\right.\right. \\
& & \hspace*{2.8cm}  \left.\left.\left. N_{\rm\bf f}(i)+J_{3}(i)N_{\rm\bf j}(i)\right]\bar{N}^{[0]}(\lambda_{0},i)
\right|_{i=1}^{N}\!\right\}
\end{eqnarray*}
Then the matrix pencil ${\Xi}_{p}^{[o]}(\lambda)$ of Theorem \ref{theorem:3} is of FCR at each $\lambda\in {\cal C}$, if and only if at each $\lambda_{0}\in{\rm\bf\Lambda}^{[0]}$, the following matrix is of FCR,
\begin{equation}
\left[\begin{array}{c}
X_{1}^{[0]}(\lambda_{0})-P_{1} Y_{1}^{[0]}(\lambda_{0})  \\
X_{2}^{[0]}(\lambda_{0})-P_{2} Y_{2}^{[0]}(\lambda_{0})  \\
X_{3}^{[0]}(\lambda_{0})-\Phi Y_{3}^{[0]}(\lambda_{0})
\end{array}\right]
\label{eqn:22}
\end{equation}
\label{theorem:5}
\end{theorem}

From the definitions of the matrices $X_{i}^{[0]}(\lambda_{0})|_{i=1}^{3}$ and $Y_{i}^{[0]}(\lambda_{0})|_{i=1}^{3}$, it is clear that each of them can be obtained through calculations with each subsystem independently. This property of the condition makes it attractive in the analysis and synthesis of a large scale NDS. On the other hand, the conditions of the above theorem can also be stated in the form of those in Theorem \ref{theorem:4}. The details are omitted due to space restrictions.

Using similar arguments as those in the proof of Theorem \ref{theorem:3}, it can be proved that the matrix pencil $\Theta(\lambda)$ of Equation (\ref{eqn:8}) is of FCR at a particular $\lambda_{0}\in{\cal C}$, if and only if the following matrix $\Theta_{p}(\lambda_{0})$ satisfies this requirement,
\begin{equation}
\Theta_{p}(\lambda_{0})\!\!=\!\!\left[\!\!\!\!\begin{array}{cccc}
\lambda_{0} E^{[0]}\!-\!A_{\rm\bf xx}^{[0]} & \lambda_{0} F_{1}\!-\!F_{2} & -A_{\rm\bf xv}^{[0]} & -J_{1} \\
-P_{1}G & M\!-\!P_{1}H & 0 & 0 \\
0 & 0 & -P_{2}K & N\!-\!P_{2}S \\
-\Phi A_{\rm\bf zx}^{[0]} & -\Phi F_{4} & I_{M_{\rm\bf v}}\!-\!\Phi A_{\rm\bf zv}^{[0]} & -\Phi J_{3} \end{array}\!\!\!\!\right]
\label{eqn:12}
\end{equation}
Moreover, the matrix ${\Xi}^{[o]}_{\infty}$ of Equation (\ref{eqn:10}) is of FCR, if and only if the matrix $\Xi^{[o]}_{\infty,p}$ defined as follows meets this condition,
\begin{equation}
\Xi_{\infty,p}^{[o]}=\left[\!\!\begin{array}{cccc}
E^{[0]} &  F_{1} & 0 & 0 \\
-C_{\rm\bf x}^{[0]} & -F_{3} & -C_{\rm\bf v}^{[0]}  & -J_{2} \\
-P_{1}G & M\!-\!P_{1}H & 0 & 0 \\
0 & 0 & -P_{2}K & N\!-\!P_{2}S \\
-\Phi A_{\rm\bf zx}^{[0]} & -\Phi F_{4} & I_{M_{\rm\bf v}}\!-\!\Phi A_{\rm\bf zv}^{[0]} & -\Phi J_{3} \end{array}\!\!\right]
\label{eqn:13}
\end{equation}
Moreover, equivalent conditions in the form of Equation (\ref{eqn:22}) can also be derived respectively for the matrix $\Theta_{p}(\lambda_{0})$/$\Xi_{\infty,p}^{[o]}$ to be of FCR. The details are omitted due to their straightforwardness.

Based on the fact that complete controllability of a descriptor system is dual to its complete observability, similar results can be obtained for NDS complete controllability verification, under the situation that the system matrices of its subsystems are expressed through some GLFTs of their (pseudo) FPPs.

On the other hand, as argued in the previous section, in order to reduce difficulties in constructing a completely observable NDS, it is helpful to select subsystems with each of them satisfying $\zeta_{Li}=0$, $i=1,2,\cdots,N$. Here, $\zeta_{Li}$ stands for the number of the matrix pencils having the form of $L_{*}(\lambda)$ in the KCF of the matrix pencil $\lambda E(i,p_{i})-\left[A_{\rm\bf xx}(i,p_{i})N_{\rm\bf x}(i,p_{i})+A_{\rm\bf xv}(i,p_{i})N_{\rm\bf v}(i,p_{i})\right]$, and its definition is given in Equation (\ref{eqn:17}). Using similar arguments as those in the proofs of Theorems \ref{theorem:1}, \ref{theorem:3} and \ref{theorem:5}, the following corollary is obtained, which provides a necessary and sufficient condition for the satisfaction of the aforementioned requirement that can be verified with each subsystem individually. Its proof is deferred to the appendix.

\begin{corollary}
Assume that the system matrices of each subsystem in the NDS $\rm\bf\Sigma$ are expressed by Equations (\ref{eqn:1-a}) and (\ref{eqn:1-b}), and FPPs of different subsystems are algebraically independent of each other. Let ${\rm\bf\Lambda}_{o}$ be a set consisting of $M_{\rm\bf x}+M_{\rm\bf v}+1$ arbitrary but distinguished complex numbers. Then $\zeta_{Li}$ of Equation (\ref{eqn:17}) is equal to zero for each $i=1,2,\cdots,N$, if and only if one of the following two conditions are satisfied,
\begin{itemize}
\setlength{\itemsep}{-0.08cm}
\item $\zeta^{[0]}_{Li}$ of Equations (\ref{eqn:20}) is equal to zero for each $i=1,2,\cdots,N$;
\item there exists a $\lambda_{0}\in {\rm\bf\Lambda}_{o}$, such that the following matrix is of FCR for each $i=1,2,\cdots,N$,
\begin{equation}
\left[\begin{array}{c}
X_{1}^{[0]}(\lambda_{0},i)-P_{1}(i) Y_{1}^{[0]}(\lambda_{0},i)  \\
X_{2}^{[0]}(\lambda_{0},i)-P_{2}(i) Y_{2}^{[0]}(\lambda_{0},i)
\end{array}\right]
\label{eqn:23}
\end{equation}
\end{itemize}
Here, $X_{j}^{[0]}(\lambda_{0},i)$ and $Y_{j}^{[0]}(\lambda_{0},i)$ stand respectively for the $i$-th diagonal block of the matrix $X_{j}^{[0]}(\lambda_{0})$ and that of the matrix $Y_{j}^{[0]}(\lambda_{0})$, in which $j=1,2$ and $i=1,2,\cdots,N$.
\label{corollary:2}
\end{corollary}

It is worthwhile to mention that before combining together to form an NDS, each subsystem usually works independently. It appears safe to declare that the assumption adopted in the above corollary is reasonable that FPPs of different subsystems are algebraically independent of each other.

Clearly, the 1st condition of Corollary \ref{corollary:2} depends only on the principles that govern the movements of a subsystem and sensor positions. That is, it is independent of any subsystem parameter. This condition is expected to be helpful in subsystem dynamics selections and sensor placements that reduce difficulties in constructing a completely observable NDS.

On the other hand, the condition of Equation (\ref{eqn:23}) provides some requirements on both the dynamics of a subsystem and its parameters. Satisfaction of this condition by each subsystem reduces significantly the number of constraints on the SCM of the NDS $\rm\bf\Sigma$, and therefore may also greatly cut down hardness in the construction of a completely observable NDS. It is believed that this condition can also provide some useful guidelines in subsystem dynamics/parameter selections.

An attractive property of the conditions in Corollary \ref{corollary:2} is that they can be verified for each subsystem individually.

Using the duality between complete observability and complete controllability of a descriptor system, similar requirements can be obtained for each subsystem with which a completely controllable NDS can be constructed more easily.

It is worthwhile to point out that in establishing results of this section and the previous section, the KCF of a matrix pencil plays an essential role. However, numerical instabilities usually arise in getting the KCF for a high dimensional matrix pencil (\cite{Duan2010,Gantmacher1959,it2017}). On the other hand, when an NDS has dense connections, the SCM $\Phi$ usually has a high dimension. These imply that when an NDS has a subsystem with many state variables and/or its subsystems are densely connected, further efforts are still required to improve numerical stability, as well as to reduce computational costs. A promising approach is to adopt matrix singular value decompositions which have been proved to be numerically quite robust.

\section{A Numerical Example}

In order to illustrate applicability of the results of the previous sections in NDS system analyses and syntheses, an artificial NDS is constructed and analyzed in this section. This NDS consists of $N$ subsystems and each of them are built by two capacitors and several  resistors. In Figure 1, a schematic illustration is given for its $i$-th subsystem with $1\leq i\leq N$, in which $x_{j}(t,i)$ with $j=1,2,3,4$ stands for the voltage of the associated node at time $t$. The first two nodes of this subsystem are connected to other subsystems, while the voltages of the remaining two nodes are measured.

\begin{figure}[!ht]
\begin{center}
\vspace{-11.0cm}\hspace*{1.8cm}\includegraphics[width=4.4in]{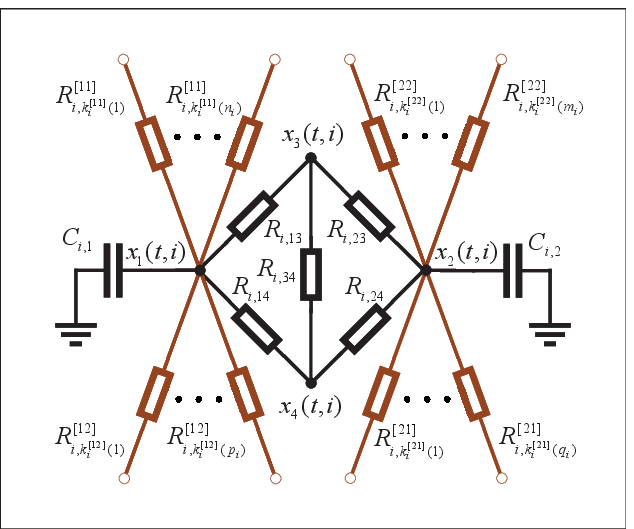}
\vspace{-0.5cm}\hspace*{5cm} \caption{The $i$-th subsystem of the NDS, in which the resistors $R_{i,k_{i}^{[\alpha\beta]}(j)}^{[\alpha\beta]}$ are used for subsystem connections.}
\end{center}
\label{figure:1}
\end{figure}
% \vspace{-0.5cm}

More specifically, the $i$-th subsystem ${\rm\bf\Sigma}_{i}$ of this artificial NDS holds the following characteristics.
\begin{itemize}
\item It has 4 nodes which are connected by resistors. Except the first two nodes, all the other nodes are directly connected, and its $\alpha$-th and $\beta$-th nodes are connected through a resistor $R_{i,\alpha\beta}$ with  $\alpha,\beta=1,2,3,4$.
\item Its $\alpha$-th node is directly connected to the $\beta$-th node of $\xi_{i}$ subsystems with their indices being $k_{i}^{[\alpha\beta]}(j)|_{j=1}^{\xi_{i}}$ and the connecting resistors equal to $R_{i,k_{i}^{[\alpha\beta]}(j)}^{[\alpha\beta]}|_{j=1}^{\xi_{i}}$, in which $\alpha,\beta\in\{1,2\}$ and $\xi\in\{m,n,p,q\}$.
\end{itemize}

Let $\xi_{i}=0$ with $\xi\in\{m,n,p,q\}$ indicate that the associated node pairs are not directly connected. Moreover, assume without any lose of generality that, for each $j_{1}, j_{2}\in\{1,2,\cdots,\xi_{i}\}$, $\alpha,\beta\in\{1,2\}$ and $\xi\in\{m,n,p,q\}$, $k_{i}^{[\alpha\beta]}(j_{1})\neq k_{i}^{[\alpha\beta]}(j_{2})$ whenever $j_{1}\neq j_{2}$. Then it is obvious that $k_{i}^{[\alpha\beta]}(j)\in \{1,2,\cdots,i-1,i+1,\cdots,N\}$ for each feasible triple $(\alpha,\beta,j)$. Moreover, $0\leq m_{i},n_{i},p_{i},q_{i}\leq N$ for each $i\in\{1,2,\cdots,N\}$.

Define the state vector $x(t,i)$ of Subsystem ${\rm\bf\Sigma}_{i}$ as
$x(t,i)={\rm\bf col}\left\{\left.x_{j}(t,i)\right|_{j=1}^{4}\right\}$. The following model can be directly established from circuit principles for its dynamics,
\begin{eqnarray*}
& &\hspace*{-1.2cm} \left[\!\!\begin{array}{cc}
C_{i} & \hspace*{-0.2cm} 0_{2\times 2}  \\
0_{2\times 2} & \hspace*{-0.2cm} 0_{2\times 2} \end{array}\!\!\right]
\!\dot{x}(t,i)\! =\!
\left[\!\!\!\begin{array}{cc}
A_{\rm\bf xx}^{[11]}(i) & A_{\rm\bf xx}^{[12]}(i)  \\
A_{\rm\bf xx}^{[21]}(i) & A_{\rm\bf xx}^{[22]}(i) \end{array}\!\!\!\right]\!
x(t,i) \!+ \\
& & \hspace*{4cm} \left[\!\!\!\begin{array}{c}
 I_{2}\\ 0_{2\times 2}\!\!\! \end{array}\right]\!\left[\!\!\!\begin{array}{c} v_{1}(t,i)\\ v_{2}(t,i) \end{array}\!\!\!\right]\\
\\
& &\hspace*{-1.2cm} {\rm\bf col}\left\{z_{1}(t,i),\; z_{2}(t,i)\right\}= \left[\begin{array}{cc} I_{2} & 0_{2\times 2} \end{array} \right]x(t,i) \\
& &\hspace*{-1.2cm} y(t,i) = \left[\begin{array}{cc}0_{2\times 2} &  I_{2}\end{array}\right]x(t,i)
\end{eqnarray*}
in which
\begin{eqnarray*}
& &\hspace*{-0.8cm}
C_{i}\!\!=\!\!\left[\!\!\begin{array}{cc}
C_{i,1} & 0  \\
0 & C_{i,2}\end{array}\!\!\right]\!\!, \hspace{0.10cm}
A_{\rm\bf xx}^{[11]}(i)\!=\!
-\!\left[\!\!\begin{array}{cc}
R_{i,13}^{-1}\!+\!R_{i,14}^{-1} & 0  \\
0 & R_{i,23}^{-1}\!+\!R_{i,24}^{-1}  \end{array}\!\!\right] \\
& &
\hspace*{-0.8cm}
A_{\rm\bf xx}^{[12]}(i)\!=\!\left[\!\begin{array}{cc}
R_{i,13}^{-1} & R_{i,14}^{-1} \\
R_{i,23}^{-1} & R_{i,24}^{-1} \end{array}\!\right], \hspace{0.15cm}
A_{\rm\bf xx}^{[21]}(i)\!=\!
\left[\!\begin{array}{cc}
R_{i,13}^{-1} & R_{i,23}^{-1}  \\
R_{i,14}^{-1} & R_{i,24}^{-1}  \end{array}\!\right] \\
& &
\hspace*{-0.8cm}
A_{\rm\bf xx}^{[22]}(i)\!=\!
-\left[\!\!\begin{array}{cc}
R_{i,13}^{-1}+R_{i,23}^{-1}+R_{i,34}^{-1} & -R_{i,34}^{-1} \\
-R_{i,34}^{-1} & R_{i,14}^{-1}+R_{i,24}^{-1}+R_{i,34}^{-1} \end{array}\!\!\right]
\end{eqnarray*}
In addition, subsystem connections are given by the following two equations,
\begin{eqnarray*}
& & \hspace*{-1.4cm} v_{1}(t,i)=\sum_{j=1}^{n_{i}}\frac{z_{1}(t,k_{i}^{[11]}(j))-z_{1}(t,i)}{R_{i,k_{i}^{[11]}(j)}^{[11]}}+ \\
& & \hspace*{2.5cm}
\sum_{j=1}^{p_{i}}\frac{z_{2}(t,k_{i}^{[12]}(j))-z_{1}(t,i)}{R_{i,k_{i}^{[12]}(j)}^{[12]}}  \\
& & \hspace*{-1.4cm} v_{2}(t,i)=\sum_{j=1}^{q_{i}}\frac{z_{1}(t,k_{i}^{[21]}(j))-z_{2}(t,i)}{R_{i,k_{i}^{[21]}(j)}^{[21]}}+\\
& & \hspace*{2.5cm}
\sum_{j=1}^{m_{i}}\frac{z_{2}(t,k_{i}^{[22]}(j))-z_{2}(t,i)}{R_{i,k_{i}^{[22]}(j)}^{[22]}}
\end{eqnarray*}

Obviously, the SCM $\Phi$ of this NDS has a dimension of $2N\times 2N$, and is completely determined by the resistors $R_{i,k_{i}^{[\alpha\beta]}(j)}^{[\alpha\beta]}|_{j=1,i=1}^{j=\xi_{i}, i=N}$ with $\alpha,\beta\in\{1,2\}$ and $\xi\in\{m,n,p,q\}$. More specifically, let $\phi_{ij}$ stand for its $i$-th row $j$-th column element. Then for each $i=1,2,\cdots,N$ and $l=1,2,\cdots,2N$, {\small
\begin{eqnarray*}
& & \hspace*{-1.0cm}
\phi_{2i-1,l}\!=\!\left\{\begin{array}{ll}
-\sum_{j=1}^{n_{i}}{1}/{R_{i,k_{i}^{[11]}(j)}^{[11]}}-\sum_{j=1}^{p_{i}}{1}/{R_{i,k_{i}^{[12]}(j)}^{[12]}}
& l=2i-1 \\
{1}/{R_{i,(l+1)/2}^{[11]}} \hspace{1.0cm} l=2k_{i}^{[11]}(j)-1,& 1\leq j\leq n_{i} \\
{1}/{R_{i,l/2}^{[12]}}\hspace{1.5cm} l=2k_{i}^{[12]}(j), & 1\leq j\leq p_{i}  \\
0 \hspace{2.5cm} otherwise & \end{array}\right. \\
& & \hspace*{-1.0cm}
\phi_{2i,l}\!=\!\left\{\begin{array}{ll}
-\sum_{j=1}^{q_{i}}{1}/{R_{i,k_{i}^{[21]}(j)}^{[21]}}-\sum_{j=1}^{m_{i}}{1}/{R_{i,k_{i}^{[22]}(j)}^{[22]}}
& l=2i \\
{1}/{R_{i,(l+1)/2}^{[21]}}\hspace{1.0cm}l=2k_{i}^{[21]}(j)-1, & 1\leq j\leq q_{i} \\
{1}/{R_{i,l/2}^{[22]}} \hspace{1.5cm}l=2k_{i}^{[22]}(j), & 1\leq j\leq m_{i} \\
0 \hspace{2.5cm} otherwise & \end{array}\right.
\end{eqnarray*}}

Note that $A_{\rm\bf zv}(i,p_{i})\equiv 0$ in this NDS. It is straightforward to prove that each of its subsystems, as well as the whole NDS, are always well-posed.

From the above subsystem dynamics/connections, it is immediate to see that the matrix pencil ${\Theta}(\lambda)$ of Equation (\ref{eqn:8}) can be expressed as follows,{\footnotesize
\begin{displaymath}
{\Theta}(\lambda)\!=\!\left[\!\!\!\!\begin{array}{cc}
{\rm\bf diag}\!\left\{\!\left.
\left[\!\!\!\begin{array}{cc}
\lambda C_{i}\!-\!A_{\rm\bf xx}^{[11]}(i) & -A_{\rm\bf xx}^{[12]}(i) \\
-A_{\rm\bf xx}^{[21]}(i) & -A_{\rm\bf xx}^{[22]}(i)  \end{array}\!\!\!\right]
\right|_{i=1}^{N} \!\right\}  &
-{\rm\bf diag}\!\left\{\!\left.\left[\!\!\!\begin{array}{c}
I_{2} \\ 0_{2\times 2} \end{array}\!\right]\right|_{i=1}^{N} \!\!\!\right\} \\
-{\Phi}\; {\rm\bf diag}\!\left\{\!\left.\left[
I_{2} \;\; 0_{2\times 2} \right]\right|_{i=1}^{N} \!\right\} & I_{2N} \end{array} \!\!\!\!\right]
\end{displaymath}}

Note that a resistor always takes a positive value. It can be directly proved from this fact that the matrix $\left[A_{\rm\bf xx}^{[21]}(i) \;\; A_{\rm\bf xx}^{[22]}(i)\right]$ is always of FCR. This implies that there exist $2\times 2$ dimensional real matrices $N_{\rm\bf xx}^{[21]}(i)$ and $N_{\rm\bf xx}^{[22]}(i)$, such that
\begin{displaymath}
\left[A_{\rm\bf xx}^{[21]}(i) \;\; A_{\rm\bf xx}^{[22]}(i)\right]^{\perp}
\!=\!\left[\begin{array}{c}
N_{\rm\bf xx}^{[21]}(i) \\
N_{\rm\bf xx}^{[22]}(i)
  \end{array}\right]
\end{displaymath}

On the basis of this equality and the consistent block diagonal structures of the submatrices in the matrix pencil ${\Theta}(\lambda)$, it can be proved using Lemma \ref{lemma:3} that the matrix pencil ${\Theta}(\lambda)$ is invertible at a particular complex value of its variable $\lambda$, if and only if the following matrix pencil holds this property,{\footnotesize
\begin{displaymath}
\left[\!\!\!\!\begin{array}{cc}
{\rm\bf diag}\!\left\{\!\left.
\lambda C_{i}N_{\rm\bf xx}^{[21]}(i)\!-\![A_{\rm\bf xx}^{[11]}(i)N_{\rm\bf xx}^{[21]}(i)\!+\!A_{\rm\bf xx}^{[12]}(i)N_{\rm\bf xx}^{[22]}(i)]
\right|_{i=1}^{N} \!\right\}  &
-I_{2N} \\
-{\Phi}\; {\rm\bf diag}\!\left\{\!\left.
N_{\rm\bf xx}^{[21]}(i)\right|_{i=1}^{N} \!\right\} & I_{2N} \end{array} \!\!\!\!\right]
\end{displaymath}}

From the well known Schur complement formula (\cite{Gantmacher1959,hj1991}), we have that the invertibility of the above matrix is equivalent to the following one,
\begin{eqnarray*}
& & \hspace*{-0.8cm} \lambda\,{\rm\bf diag}\!\left\{\!\left.
 C_{i}N_{\rm\bf xx}^{[21]}(i)\right|_{i=1}^{N} \!\right\}-{\Phi}\, {\rm\bf diag}\!\left\{\!\left.
N_{\rm\bf xx}^{[21]}(i)\right|_{i=1}^{N} \!\right\} \\
& & \hspace*{1.1cm}
-{\rm\bf diag}\!\left\{\!\left.
 [A_{\rm\bf xx}^{[11]}(i)N_{\rm\bf xx}^{[21]}(i)\!+\!A_{\rm\bf xx}^{[12]}(i)N_{\rm\bf xx}^{[22]}(i)]
\right|_{i=1}^{N} \!\right\}
\end{eqnarray*}

Note that both a capacitor and a resistor can take only a positive value, which implies that for each $i=1,2,\cdots,N$, the matrix $C_{i}N_{\rm\bf xx}^{[21]}(i)$ can not equal to zero. It can therefore be declared that there certainly exists a $\lambda_{0}\in{\cal C}$, such that the value of the last matrix pencil is invertible. That is, the matrix pencil $\Theta(\lambda)$ is of FNCR. Hence, it can be claimed from Theorem \ref{theorem:1} that the NDS $\rm\bf\Sigma$ is regular.

On the other hand, it can also be easily seen from the subsystem dynamics and connections that, the matrix ${\Xi}_{\infty}^{[o]}$ of Equation (\ref{eqn:10}) can be expressed as
\begin{equation}
{\Xi}_{\infty}^{[o]}= \left[\!\begin{array}{cc}
{\rm\bf diag}\!\left\{\!\left.\left[\!\begin{array}{cc}
C_{i} & 0_{2\times 2} \\
0_{2\times 2} & 0_{2\times 2} \end{array}\!\right]\right|_{i=1}^{N} \!\right\} & 0 \\
{\rm\bf diag}\!\left\{\!\left.\left[\!\begin{array}{cc}
0_{2\times 2} & -I_{2} \end{array}\!\right]\right|_{i=1}^{N} \!\right\} & 0 \\
-\Phi\;{\rm\bf diag}\!\left\{\!\left.\left[\!\begin{array}{cc}
I_{2} & 0_{2\times 2} \end{array}\!\right]\right|_{i=1}^{N} \!\right\} & I_{2N}
\end{array}\!\right]
\label{eqn:24}
\end{equation}
which is obviously of FCR for an arbitrary SCM $\Phi$. On the other hand,
direct algebraic manipulations show that for every $i\in\{1,2,\cdots,N\}$, we have that
\begin{equation}
\hspace*{-0.0cm} {\rm\bf null}\!\left(\left[ -C_{\rm\bf x}(i,p_{i})\;\;\; -C_{\rm\bf v}(i,p_{i})\right]\right) \!=\!
{\rm\bf span}\!\left(\!\left[\!\begin{array}{cc}
I_{2} & 0_{2\times 2} \\
0_{2\times 2} & 0_{2\times 2}  \\
0_{2\times 2} & I_{2}  \end{array}\!\right]\!\right)
\label{eqn:28}
\end{equation}
Therefore,
\begin{eqnarray}
& & \hspace*{-0.6cm}\left[\!\!\begin{array}{cc}
\lambda E(p)\!-\!A_{\rm\bf xx}(p) & -A_{\rm\bf xv}(p) \\
-\Phi A_{\rm\bf zx}(p) & I_{2N}\!-\!\Phi\, A_{\rm\bf zv}(p) \end{array}\!\!\right]\!\left[-C_{\rm\bf x}(p) \;\; -C_{\rm\bf v}(p)\right]^{\perp} \nonumber\\
& &\hspace*{-1.0cm}=
\left[\!\!\begin{array}{c}
{\rm\bf diag}\!\left\{\!\left.\left[\!\!\!\begin{array}{cc}
\lambda C_{i}\!-\!A^{[11]}_{\rm\bf xx}(i) & -I_{2} \\
-A^{[21]}_{\rm\bf xx}(i) & 0_{2\times 2} \end{array}\!\!\!\right]\right|_{i=1}^{N} \right\}  \\
{\rm\bf diag}\!\left\{\!\left.\left[0_{2\times 2}\;\; I_{2}\right]\right|_{i=1}^{N}\!\right\} - \Phi\, {\rm\bf diag}\!\left\{\!\left.\left[I_{2}\;\; 0_{2\times 2}\right]\right|_{i=1}^{N} \!\right\}
\end{array}\!\!\right]
\label{eqn:25}
\end{eqnarray}

On the basis of this equality and Lemma \ref{lemma:3}, it can be declared that if for each $i=1,2,\cdots,N$, $ R_{i,13}R_{i,24}\neq R_{i,23}R_{i,14}$, that is, the matrix $A^{[21]}_{\rm\bf xx}(i)$ is invertible, then the  matrix pencil ${\Xi}^{[o]}(\lambda)$ of Equation (\ref{eqn:9}) is of FCR at each $\lambda\in{\cal C}$,
no matter how many subsystems the NDS has, how its subsystems are connected, and what values the capacitors and resistors take.

From Theorem \ref{theorem:2}, these results means that as long as $ R_{i,13}R_{i,24}\neq R_{i,23}R_{i,14}$ is satisfied by each subsystem individually, the artificial NDS is always completely observable.

When there is an $i\in\{1,2,\cdots,N\}$ such that the associated $A^{[21]}_{\rm\bf xx}(i)$ is singular, straightforward matrix operations show that
\begin{equation}
\left[\!\begin{array}{cc}
-A^{[21]}_{\rm\bf xx}(i) & 0_{2\times 2} \end{array}\!\right]^{\perp}=
\left[\!\begin{array}{cc}
R_{i,13} & 0_{1\times 2} \\
-R_{i,23} & 0_{1\times 2} \\
0_{2\times 1}  & I_{2} \end{array}\!\right]
\label{eqn:26}
\end{equation}
With this equality and Lemma \ref{lemma:3}, as well as Theorem \ref{theorem:4}, necessary and sufficient conditions can be straightforwardly derived about the SCM $\Phi$ for the NDS complete observability.

More specifically, let ${\cal I\!}_{s}$ denote the set of subsystem indices with the associated matrix $A^{[21]}_{\rm\bf xx}(i)$ not invertible. Then from Lemma \ref{lemma:3} and Equation (\ref{eqn:26}), we have that the matrix pencil on the right hand side of Equation (\ref{eqn:25}) is of FCR at each $\lambda\in{\cal C}$, if and only if the following matrix pencil is, {\footnotesize
\begin{displaymath}
\left[\!\!\!\begin{array}{c}
{\rm\bf diag}\!\left\{\left.\left[\!\!\!\begin{array}{ccc}
\lambda R_{i,13}C_{i,1}\!+\!1\!+\!R_{i,13}/R_{i,14} & -1 & 0 \\
-\lambda R_{i,23}C_{i,2}\!-\!1-\!R_{i,13}/R_{i,14}  & 0 & -1 \end{array}\!\!\!\right]\right|_{i\in {\cal I\!}_{s}} \right\}  \\
{\rm\bf diag}\!\left\{\!\left.\left[\!\!\!\begin{array}{c}
0 \\
{[0_{2\times 1} \;\; I_{2}]}
\\ 0
\end{array}\!\!\!\right]\right|_{i\in {\cal I\!}_{s}}\right\} \!-\! \Phi\, {\rm\bf diag}\!\left\{\!\!\!\left.\left[\!\!\!\!\begin{array}{c}
0 \\
\left[\!\!\begin{array}{cc} R_{i,13} & 0_{1\times 2} \\ -R_{i,23} & 0_{1\times 2}\end{array}\!\!\!\!\right]\\
0
\end{array}\!\!\right]
\right|_{i\in {\cal I\!}_{s}} \right\}
\end{array}\!\!\!\!\right]
\end{displaymath}}
On the basis of Theorem \ref{theorem:4}, it can be further proved that under such a situation, the artificial NDS $\rm\bf\Sigma$ is completely observable, if and only if the following matrix pencil is of FCR at each $\lambda\in{\cal C}$,
\begin{eqnarray*}
& &\!\!\!\! {\rm\bf diag}\!\left\{\!\left.\left[\!\!\!\begin{array}{c}
0 \\
\lambda R_{i,13}C_{i,1}\!+\!1\!+\!R_{i,13}/R_{i,14} \\
-\lambda R_{i,23}C_{i,2}\!-\!1-\!R_{i,13}/R_{i,14}  \\
0\end{array}\!\!\!\right]\right|_{i\in {\cal I\!}_{s}} \right\} \\
& -& \!\!\!\!
\Phi\, {\rm\bf diag}\!\left\{\!\left.\left[\!\!\!\begin{array}{c} 0 \\ R_{i,13}  \\ -R_{i,23}\\ 0 \end{array}\!\!\!\right]\right|_{i\in {\cal I\!}_{s}} \right\}
\end{eqnarray*}

Verification of this condition is possible when the element number of the set ${\cal I\!}_{s}$ is moderate. When this number is large, further efforts are still required for developing computationally attractive verification methods.

\section{Concluding Remarks}

This paper investigates regularity, complete controllability and complete observability for a networked dynamic system, in which each subsystem is described by a descriptor form, while its system matrices is represented by a generalized linear fractional transformation of its (pseudo) first principle parameters. Some matrix rank based necessary and sufficient conditions have been derived, in which the matrix depends affinely on both the subsystem connection matrix and subsystem parameters. These results extends those on controllability/observability of an NDS with its subsystems being described by a state space model, and keep the attractive properties that all the involved calculations can be performed on each subsystem independently. In addition, these conditions also reveals some requirements on a subsystem from which a completely controllable/observable NDS can be constructed more easily, which are expected to be helpful in subsystem designs and parameter selections.

As a further issue, it is interesting to see applicability of the obtained results to actuator/sensor placements for an NDS, as well as to develop quantitative measures for its controllability/observability. Another important issue is to see possibilities of extending these results to causality verifications for this kind of NDSs.

\vspace{0.25cm}
\hspace*{-0.45cm}{\rm\bf Acknowledgements.}
The author would like to thank Mr. Y. Y. Zhou for his help and efforts in constructing the numerical example.

\renewcommand{\theequation}{a.\arabic{equation}}
\setcounter{equation}{0}

% \vspace{-0.25cm}
\section*{Appendix: Proof of Some Technical Results}
% \vspace{-0.25cm}

\hspace*{-0.45cm}{\rm\bf Proof of Theorem 1:} When both the subsystems and the whole system of the NDS ${\rm\bf\Sigma}$ are well-posed, direct matrix manipulations show that the matrix $I_{M_{\rm\bf v}}-\Phi  A_{\rm\bf zv}(p)$ is invertible (\cite{Zhou2015,zyl2018}). Hence
\begin{eqnarray}
{\rm\bf det}\left\{\Theta(\lambda)\right\}\! \!\!\!\!\!
&=&\!\!\!\!\!\! {\rm\bf det}\left(I_{M_{\rm\bf v}}\!\!-\!\Phi  A_{\rm\bf zv}(p) \right)\!{\rm\bf det}\!\left(\! \lambda E(p) \!-\!\left[ \!A_{\rm\bf xx}(p)  \right.\right. \nonumber \\
& & \hspace*{0.2cm}
 \left.\left. + A_{\rm\bf xv}(p)\!\left[\! I_{M_{\rm\bf v}} \!-\! \Phi  A_{\rm\bf zv}(p)\!\right]^{-1}\!\!\Phi  A_{\rm\bf zx}(p)\!\right] \right) \nonumber\\
&=&\!\!\!\!\!\! {\rm\bf det}\!\left(I_{M_{\rm\bf v}}\!-\!\Phi  A_{\rm\bf zv}(p)   \right)\!\times\!
{\rm\bf det}\!\left( \lambda E(p) - A(p)\right) \nonumber\\
\label{eqn:a-1}
\end{eqnarray}

This means that at each $\lambda$, the nonsingularity of the matrix pencil $\Theta(\lambda)$ is equal to that of the matrix pencil $\lambda E(p) -A(p)$.

Assume now that there is a $\lambda_{0}$ belonging to the set ${\rm\bf \Lambda}_{r}$ such that the matrix $\Theta(\lambda_{0})$ is of FCR. Then, the above arguments means that at this particular value, the matrix pencil $\lambda E(p)-A(p)$ is invertible. The regularity of the NDS ${\rm\bf\Sigma}$ follows from its definition for a descriptor system.

On the other hand, assume that for each $\lambda_{0}\in{\rm\bf \Lambda}_{r}$, the matrix pencil $\Theta(\lambda)$ is rank deficient. Then the equivalence in the nonsingularity between the matrix $\Theta(\lambda_{0})$ and the matrix $\lambda_{0} E(p) -A(p)$ implies that ${\rm\bf det}\left[ \lambda_{0} E(p) -A(p)\right]=0$ whenever $\lambda_{0}$ is an element of the set ${\rm\bf \Lambda}_{r}$. On the other hand, note that ${\rm\bf det}\left[ \lambda E(p) -A(p)\right]$ is a polynomial of the variable $\lambda$ with its degree not exceeding $M_{\rm\bf x}$. This means that if ${\rm\bf det}\left[ \lambda E(p) -A(p)\right]$ is not a zero polynomial, it has at most $M_{\rm\bf x}$ roots. In addition, recall that the set ${\rm\bf \Lambda}_{r}$ consists of $M_{\rm\bf x}+1$ different elements. It can therefore be declared that ${\rm\bf det}\left[\Theta(\lambda_{0})\right]=0$ for each $\lambda_{0}\in{\rm\bf \Lambda}_{r}$ means that ${\rm\bf det}\left[ \lambda E(p) -A(p)\right]\equiv 0$. Hence, the NDS ${\rm\bf\Sigma}$ is not regular, according to the definition of a descriptor system.

The proof is now completed.    \hspace{\fill}$\Diamond$

\hspace*{-0.45cm}{\rm\bf Proof of Theorem 2:} To shorten mathematical expressions in this proof, the dependence of a matrix on the parameter vector $p$ is eliminated for each associated matrix. For example, the matrix $A_{\rm\bf xx}(p)$ is written as $A_{\rm\bf xx}$, etc. This elimination does not introduce any confusions in the following derivations.

Assume that the NDS ${\rm\bf\Sigma}$ is completely observable. Then according to Lemma \ref{lemma:2} and Equation (\ref{eqn:7}), we have that for every $\lambda\in {\cal C}$ and every nonzero $M_{\rm\bf x}$ dimensional complex vector $\alpha$,
\begin{equation}
\left[\begin{array}{c} \lambda E -\left[A_{\rm\bf xx}+A_{\rm\bf xv}
(I_{M_{\rm\bf v}}-\Phi A_{\rm\bf zv})^{-1}\Phi A_{\rm\bf zx}\right] \\
C_{\rm\bf x}+C_{\rm\bf v}
(I_{M_{\rm\bf v}}-\Phi  A_{\rm\bf zv})^{-1}\Phi A_{\rm\bf zx}\end{array}\right]\alpha \neq 0 \label{eqn:a6}
\end{equation}

Now, assume that at a particular $\lambda_{0}$, the matrix pencil ${\Xi}^{[o]}(\lambda)$ is not of FCR. Then a nonzero $M_{\rm\bf x}+M_{\rm\bf v}$ dimensional complex vector $\xi$ exists  satisfying
\begin{equation}
{\Xi}^{[o]}(\lambda_{0})\xi=0 \label{eqn:a1}
\end{equation}

Partition the vector $\xi$ as $\xi={\rm\bf col}\{\xi_{1},\;\xi_{2}\}$ consistently with the matrix pencil ${\Xi}^{[o]}(\lambda)$. Then Equation (\ref{eqn:a1}) can be equivalently rewritten as
\begin{eqnarray}
& & (\lambda_{0} E -A_{\rm\bf xx})\xi_{1}-A_{\rm\bf xv}\xi_{2}=0 \label{eqn:a2}\\
& & C_{\rm\bf x}\xi_{1}+C_{\rm\bf v}\xi_{2}=0 \label{eqn:a3} \\
& & -\Phi  A_{\rm\bf zx}\xi_{1}+(I_{M_{\rm\bf v}}-\Phi  A_{\rm\bf zv})\xi_{2}=0
\label{eqn:a4}
\end{eqnarray}

On the other hand, from the well-posedness assumptions on each subsystem and the whole system of the NDS ${\rm\bf\Sigma}$, it can be directly proved that the matrix $I_{M_{\rm\bf v}}-\Phi  A_{\rm\bf zv}$ is invertible (\cite{Zhou2015,zyl2018}). It can therefore be declared from Equation (\ref{eqn:a4}) and the assumption $\xi\neq 0$ that $\xi_{1}\neq 0$ and
\begin{equation}
\xi_{2}=(I_{M_{\rm\bf v}}-\Phi  A_{\rm\bf zv})^{-1}\Phi  A_{\rm\bf zx}\xi_{1}
\label{eqn:a5}
\end{equation}

Combing Equations (\ref{eqn:a2}), (\ref{eqn:a3}) and (\ref{eqn:a5}) together, we further have that
\begin{equation}
\hspace*{-0.25cm} \left[\!\!\begin{array}{c}
\lambda_{0} E \!-\! \left[A_{\rm\bf xx}+A_{\rm\bf xv}
(I_{M_{\rm\bf v}} \!-\! \Phi  A_{\rm\bf zv})^{-1}\Phi  A_{\rm\bf zx}\right] \\
C_{\rm\bf x}+C_{\rm\bf v}
(I_{M_{\rm\bf v}}-\Phi  A_{\rm\bf zv})^{-1}\Phi  A_{\rm\bf zx}\end{array} \!\!\right]\!\xi_{1} \!=\! 0
\end{equation}
which is clearly in contradiction with Equation (\ref{eqn:a6}). Hence, the matrix pencil ${\Xi}^{[o]}(\lambda)$ must be  of FCR at each complex number $\lambda$.

Now assume that the matrix ${\Xi}_{\infty}$ is not of FCR. Similar arguments show that it will lead to the rank deficiency of the matrix ${\rm\bf col}\{E,\; C\}$, which further results that the NDS ${\rm\bf\Sigma}$ is not completely observable.

On the contrary, assume that the NDS ${\rm\bf\Sigma}$ is not completely observable. Then according to Lemma \ref{lemma:2}, there exist a $\lambda_{0}\in {\cal C}$ and a nonzero vector $\alpha$ such that
\begin{equation}
\hspace*{-0.25cm} \left[\!\!\begin{array}{c}
\lambda_{0} E \!-\! \left[A_{\rm\bf xx}+A_{\rm\bf xv}
(I_{M_{\rm\bf v}} \!-\! \Phi  A_{\rm\bf zv})^{-1}\Phi  A_{\rm\bf zx}\right] \\
C_{\rm\bf x}+C_{\rm\bf v}
(I_{M_{\rm\bf v}}-\Phi  A_{\rm\bf zv})^{-1}\Phi  A_{\rm\bf zx}\end{array} \!\!\right]\!\alpha \!=\! 0
\label{eqn:a8-1}
\end{equation}
or there is a nonzero vector $\alpha$ such that
\begin{equation}
\hspace*{-0.25cm} \left[\!\!\begin{array}{c}
E \\
C_{\rm\bf x}+C_{\rm\bf v}
(I_{M_{\rm\bf v}}-\Phi  A_{\rm\bf zv})^{-1}\Phi  A_{\rm\bf zx}\end{array} \!\!\right]\!\alpha \!=\! 0
\label{eqn:a8-2}
\end{equation}

Assume now that Equation (\ref{eqn:a8-1}) is satisfied. Define a vector $\xi$ as
\begin{displaymath}
\xi=\left[\begin{array}{c}
I_{M_{\rm\bf x}} \\
(I_{M_{\rm\bf v}}-\Phi  A_{\rm\bf zv})^{-1}\Phi  A_{\rm\bf zx}
 \end{array}\right] \alpha
\end{displaymath}
Then $\xi\neq 0$ and the following equality is obviously satisfied by this vector,
\begin{equation}
\hspace*{-0.25cm} \left[
-\Phi  A_{\rm\bf zx} \;\; I_{M_{\rm\bf v}}-\Phi  A_{\rm\bf zv}\right]\!\xi = 0
\label{eqn:a8-3}
\end{equation}
Moreover, Equation (\ref{eqn:a8-1}) can be equivalently rewritten as
\begin{equation}
\hspace*{-0.25cm} \left[\!\!\begin{array}{cc}
\lambda_{0} E \!-\! A_{\rm\bf xx} & -A_{\rm\bf xv} \\
-C_{\rm\bf x} & -C_{\rm\bf v}\end{array} \!\!\right]\!\xi \!=\! 0
\label{eqn:a8-4}
\end{equation}
It can therefore be declared from the definition of the matrix pencil ${\Xi}^{[o]}(\lambda)$ that
\begin{equation}
{\Xi}^{[o]}(\lambda_{0})\xi =0 \label{eqn:a9}
\end{equation}
That is, this matrix pencil is not of FCR for each $\lambda\in {\cal C}$.

Now assume that Equation (\ref{eqn:a8-2}) is satisfied. Similar arguments as those for the situation in which Equation (\ref{eqn:a8-1}) is satisfied show that, there exists a nonzero vector $\xi$ satisfying
\begin{equation}
\hspace*{-0.25cm} \left[\!\!\begin{array}{cc}
E  & 0 \\
-C_{\rm\bf x} & -C_{\rm\bf v}   \\
-\Phi  A_{\rm\bf zx} & I_{M_{\rm\bf v}}-\Phi  A_{\rm\bf zv}
\end{array}\!\!\right]\!\xi \!=\! 0
\end{equation}

This completes the proof.   \hspace{\fill}$\Diamond$

\hspace*{-0.45cm}{\rm\bf Proof of Theorem 3:} For brevity, denote the following matrix pencils with $1\leq i\leq N$
\begin{eqnarray*}
& & \hspace*{0.5cm}
\lambda E(i)N_{\rm\bf x}(i)-\left[A_{\rm\bf xx}(i)N_{\rm\bf x}(i)+A_{\rm\bf xv}(i)N_{\rm\bf v}(i)\right] \\
& & \hspace*{-0.6cm}
{\rm\bf diag}\!\left\{\!H_{\xi_{Hi}}(\lambda),\;K_{\xi_{Ki}(j)}(\lambda)|_{j=1}^{\zeta_{Ki}},\; L_{\xi_{Li}(j)}(\lambda)|_{j=1}^{\zeta_{Li}},\right.\nonumber \\
& & \hspace*{3.5cm}\left.  N_{\xi_{Ni}(j)}(\lambda)|_{j=1}^{\zeta_{Ni}},\; J_{\xi_{Ji}(j)}(\lambda)|_{j=1}^{\zeta_{Ji}}\!\right\}
\end{eqnarray*}
and
\begin{displaymath}
{\rm\bf diag}\!\left\{\!H_{\xi_{Hi}}(\lambda),\;K_{\xi_{Ki}(j)}(\lambda)|_{j=1}^{\zeta_{Ki}},\; L_{\xi_{Li}(j)}(\lambda)|_{j=1}^{\zeta_{Li}}\right\}
\end{displaymath}
respectively by $\Pi(\lambda,i)$, $\bar{\Psi}(\lambda,i)$ and $\Psi(\lambda,i)$.

Let $\alpha$ be an arbitrary $M_{\rm\bf x}+M_{\rm\bf v}$ dimensional real column vector. Partition this vector as $\alpha={\rm\bf col}\!\left\{\alpha_{\rm\bf x}(i)|_{i=1}^{N},\;\alpha_{\rm\bf v}(i)|_{i=1}^{N}\right\}$, in which $\alpha_{{\rm\bf x}}(i)$ belongs to ${\cal C}^{m_{{\rm\bf x}i}}$, while $\alpha_{{\rm\bf v}}(i)$ belongs to ${\cal C}^{m_{{\rm\bf v}i}}$, $i=1,2,\cdots,N$. From the block diagonal structure of the matrix $C_{\rm\bf x}$ and $C_{\rm\bf v}$, it is immediate that
\begin{equation}
\left[ C_{\rm\bf x}\;\; C_{\rm\bf v} \right]\alpha
=\left[\begin{array}{c}
C_{\rm\bf x}(1)\alpha_{{\rm\bf x}}(1)+ C_{\rm\bf v}(1)\alpha_{{\rm\bf v}}(1) \\
C_{\rm\bf x}(2)\alpha_{{\rm\bf x}}(2)+ C_{\rm\bf v}(2)\alpha_{{\rm\bf v}}(2) \\
\vdots \\
C_{\rm\bf x}(N)\alpha_{{\rm\bf x}}(N)+ C_{\rm\bf v}(N)\alpha_{{\rm\bf v}}(N) \end{array}\right]
\end{equation}

From this relation and the definitions of the matrices $N_{\rm\bf x}(i)$ and $N_{\rm\bf v}(i)$, as well as Lemma \ref{lemma:4}, direct algebraic operations show that
\begin{equation}
{\rm\bf Null}\left([C_{\rm\bf x} \;\;C_{\rm\bf v}]\right)
={\rm\bf Span}\left(\left[\begin{array}{c} N_{\rm\bf x}  \\ N_{\rm\bf v} \end{array}\right]\right)
\label{eqn:a15}
\end{equation}
in which
\begin{displaymath}
N_{\rm\bf x}={\rm\bf diag}\left\{\left. N_{\rm\bf x}(i)\right|_{i=1}^{N}\right\},\hspace{0.5cm}
N_{\rm\bf v}={\rm\bf diag}\left\{\left. N_{\rm\bf v}(i)\right|_{i=1}^{N}\right\}
\end{displaymath}

On the basis of Equation (\ref{eqn:a15}) and Lemma \ref{lemma:3}, it is clear that at an arbitrary $\lambda_{0}\in{\cal C}$, the value of the matrix pencil ${\Xi}^{[o]}(\lambda)$ which is defined in Equation (\ref{eqn:9}), that is, the matrix ${\Xi}^{[o]}(\lambda_{0})$, is of FCR, if and only if the following matrix is,
\begin{eqnarray}
& &
\left[\begin{array}{cc}
\lambda_{0} E-A_{\rm\bf xx} & -A_{\rm\bf xv} \\
-\Phi A_{\rm\bf zx} & I_{M_{\rm\bf v}}-\Phi A_{\rm\bf zv} \end{array}\right]\left[\begin{array}{c} N_{\rm\bf x}  \\ N_{\rm\bf v} \end{array}\right]\nonumber\\
&=&
\left[\begin{array}{c}
\lambda_{0} E N_{\rm\bf x}-\left[A_{\rm\bf xx}N_{\rm\bf x}+A_{\rm\bf xv}N_{\rm\bf v}\right] \\
N_{\rm\bf v}-\Phi \left[A_{\rm\bf zx}N_{\rm\bf x}+A_{\rm\bf zv}N_{\rm\bf v}\right] \end{array}\right]
\label{eqn:a16}
\end{eqnarray}

From the consistent block diagonal structures of the involved matrices, as well as Equation (\ref{eqn:17}), we have that
\begin{eqnarray}
& & \hspace*{-1.0cm}
\lambda_{0} E N_{\rm\bf x}-\left[A_{\rm\bf xx}N_{\rm\bf x}+A_{\rm\bf xv}N_{\rm\bf v}\right]   \nonumber\\
& &\hspace*{-1.2cm}=\! {\rm\bf diag}\left\{\left. \Pi(\lambda_{0},i)\right|_{i=1}^{N}\right\}  \nonumber\\
& &\hspace*{-1.2cm}=\! {\rm\bf diag}\!\left\{\!\!\left. U(i)\right|_{i=1}^{N}\!\!\right\}
\!{\rm\bf diag}\!\left\{\!\!\left.\bar{\Psi}(\lambda_{0},i)\right|_{i=1}^{N}\!\!\right\}\!{\rm\bf diag}\!\left\{\!\!\left. V(i)\right|_{i=1}^{N}\!\!\right\}
\label{eqn:a17}
\end{eqnarray}
Recall that for each $i=1,2,\cdots,N$, both the matrix $U(i)$ and the matrix $V(i)$ are invertible. We therefore have that,
\begin{eqnarray}
& & \left[\begin{array}{c}
\lambda_{0} E N_{\rm\bf x}-\left[A_{\rm\bf xx}N_{\rm\bf x}+A_{\rm\bf xv}N_{\rm\bf v}\right] \\
N_{\rm\bf v}-\Phi \left[A_{\rm\bf zx}N_{\rm\bf x}+A_{\rm\bf zv}N_{\rm\bf v}\right] \end{array}\right] \nonumber\\
&=& \left[\!\!\!\begin{array}{cc}
{\rm\bf diag}\left\{\!\left. U(i)\right|_{i=1}^{N}\right\} & 0 \\
0 & I \end{array} \!\!\right]\!\times   \nonumber\\
& & \left[\!\!\!\!\begin{array}{c}
{\rm\bf diag}\left\{\left.\bar{\Psi}(\lambda_{0},i)\right|_{i=1}^{N}\right\} \\
\left[N_{\rm\bf v}\!-\!\Phi \left(A_{\rm\bf zx}N_{\rm\bf x}\!+\!A_{\rm\bf zv}N_{\rm\bf v}\right)\right]{\rm\bf diag}\!\left\{\!\left. V^{-1}(i)\right|_{i=1}^{N}\right\}
\end{array}\!\!\!\!\right]\!\times \nonumber\\
& &
 \hspace*{4.2cm}{\rm\bf diag}\!\left\{\!\left. V(i)\right|_{i=1}^{N}\right\}
\label{eqn:a23}
\end{eqnarray}
Hence, it can be declared from the definitions of the matrix pencils $\bar{\Psi}(\lambda,i)|_{i=1}^{N}$ that, the matrix pencil ${\Xi}^{[o]}(\lambda)$ is always of FCR, only if the following matrix pencil $\hat{\Xi}^{[o]}(\lambda)$ is
\begin{equation}
\hat{\Xi}^{[o]}(\lambda)\!=\!\left[\!\!\!\!\!\!\!\begin{array}{c}
{\rm\bf diag}\!\left\{\! \left.{\rm\bf diag}\!\left\{ \!\left. L_{\xi_{Li}(j)}(\lambda)\right|_{j=1}^{\zeta_{Li}}\right\}\right|_{i=1}^{N} \!\right\}
\\
\hspace*{-0.7cm}\left[N_{\rm\bf v}\!-\!\Phi \left(A_{\rm\bf zx}N_{\rm\bf x}\!+\!A_{\rm\bf zv}N_{\rm\bf v}\right)\right]\!\times \\
\hspace*{1.5cm}{\rm\bf diag}\!\left\{\!\left. V^{-1}_{i}(n(i):m(i))\right|_{i=1}^{N}\right\}
\end{array}\!\!\!\!\right]
\label{eqn:a24}
\end{equation}
Moreover, using Lemmas \ref{lemma:0} and \ref{lemma:3}, it can be straightforwardly shown that the latter is equivalent to that the MVP $\Omega(\lambda)-\Phi\Gamma(\lambda)$ is always of FCR.

On the other hand, recall that a matrix pencil in the form of $N_{*}(\lambda)$ or $J_{*}(\lambda)$ is always of FCR. Then Equation (\ref{eqn:a23}) and repetitive applications of Lemma \ref{lemma:4} lead to that the matrix ${\Xi}^{[o]}(\lambda_{0})$ is of FCR, if and only if the following matrix is of FCR,
\begin{equation}
\left[\!\!\begin{array}{c}
{\rm\bf diag}\left\{\left.{\Psi}(\lambda_{0},i)\right|_{i=1}^{N}\right\} \\
\left[N_{\rm\bf v}\!-\!\Phi \left(A_{\rm\bf zx}N_{\rm\bf x}\!+\!A_{\rm\bf zv}N_{\rm\bf v}\right)\right]{\rm\bf diag}\left\{\left. V_{i}^{-1}(m(i))\right|_{i=1}^{N}\right\}
\end{array}\!\!\right]
\label{eqn:a19-a}
\end{equation}

Assume now that $\lambda_{0}\in{\rm\bf\Lambda}$. The definition of the matrix $N(\lambda_{0},i)$ and the block diagonal structure of the associated matrix imply that
\begin{equation}
{\rm\bf null}\left(\!{\rm\bf diag}\left\{\!\left.{\Psi}(\lambda_{0},i)\right|_{i=1}^{N}\right\}\!\right)
\!=\!{\rm\bf span}\left(\!{\rm\bf diag}\left\{\left. N(\lambda_{0},i)\right|_{i=1}^{N}\right\}\!
\right)
\label{eqn:a20}
\end{equation}
In addition, it can be directly declared from Lemma \ref{lemma:4} that the matrix ${\rm\bf diag}\left\{\left. N(\lambda_{0},i)\right|_{i=1}^{N}\right\}$ is of FCR. Based on this relation and Lemma \ref{lemma:3}, we have that the matrix of Equation (\ref{eqn:a19-a}) is of FCR, if and only if the following matrix is of FCR,
\begin{eqnarray*}
& & \hspace*{-1.0cm}\left[N_{\rm\bf v}\!-\!\Phi \left(A_{\rm\bf zx}N_{\rm\bf x}\!+\!A_{\rm\bf zv}N_{\rm\bf v}\right)\right]{\rm\bf diag}\!\left\{\left.\! V_{i}^{-1}(m(i))\right|_{i=1}^{N}\!\right\}\times \\
& & \hspace*{4.2cm} {\rm\bf diag}\!\left\{\left.\! N(\lambda_{0},i)\right|_{i=1}^{N}\!\right\}
\end{eqnarray*}
which is obviously equal to the matrix $X(\lambda_{0})-\Phi Y(\lambda_{0})$.

Now assume that $\lambda_{0}\in{\cal C}\backslash{\rm\bf\Lambda}$. According to the definitions of the set ${\rm\bf\Lambda}$, we have that for each
$1\leq j\leq \zeta_{Ki}$ and each $1\leq i\leq N$, both the matrix $H_{\xi_{Hi}}(\lambda_{0})$ and the matrix $K_{\xi_{Ki}(j)}(\lambda_{0})$ are of FCR. Hence, Lemma \ref{lemma:4} leads to that the matrix of Equation (\ref{eqn:a19-a}) is of FCR, if and only if the matrix $\hat{\Xi}^{[o]}(\lambda_{0})$ is. The latter is obviously guaranteed from the arguments immediately after Equation (\ref{eqn:a24}), provided that the 2nd condition of the theorem is satisfied. This completes the proof.   \hspace{\fill}$\Diamond$

\hspace*{-0.45cm}{\rm\bf Proof of Theorem 4:} Assume that at a particular $\lambda_{0}\in{\cal C}$, the matrix pencil ${\Xi}^{[o]}(\lambda)$ is rank deficient. Then, there exist vectors $\alpha$ and $\beta$ such that at least one of them is nonzero and the following equation is satisfied,
\begin{equation}
\left[\begin{array}{cc}
\lambda_{0} E(p)-A_{\rm\bf xx}(p) & -A_{\rm\bf xv}(p) \\
-C_{\rm\bf x}(p) & -C_{\rm\bf v}(p) \\
-\Phi A_{\rm\bf zx}(p) & I_{M_{\rm\bf z}}-\Phi A_{\rm\bf zv}(p) \end{array}\right]
\left[\begin{array}{c} \alpha \\ \beta \end{array}\right]
=0
\label{eqn:a10}
\end{equation}

Define  vectors $\xi$ and $\eta$ respectively as
\begin{displaymath}
\xi=(M-P_{1}H)^{-1}P_{1}G\alpha, \hspace{0.5cm} \eta=(N-P_{2}S)^{-1}P_{2}K\beta
\end{displaymath}
Then, the vectors $\alpha$, $\beta$, $\xi$ and $\eta$ obviously satisfy
\begin{eqnarray}
& & \left[-P_{1}G\;\; M-P_{1}H\right]\left[\begin{array}{c} \alpha \\ \xi \end{array}\right]=0
\label{eqn:a11}  \\
& & \left[-P_{2}K\;\; N-P_{2}S\right]\left[\begin{array}{c} \beta \\ \eta \end{array}\right]=0
\label{eqn:a12}
\end{eqnarray}

On the other hand, from the definitions of the associated matrices which are given by Equations (\ref{eqn:1-a}) and (\ref{eqn:1-b}), as well as the paragraph immediately before this theorem, Equation (\ref{eqn:a10}) can be rewritten as
\begin{equation}
\left[\!\!\begin{array}{cccc}
\lambda_{0} E^{[0]}\!-\!A_{\rm\bf xx}^{[0]} & \lambda_{0} F_{1}\!-\!F_{2} & -A_{\rm\bf xv}^{[0]} & -J_{1} \\
-C_{\rm\bf x}^{[0]} & -F_{3} & -C_{\rm\bf v}^{[0]}  & -J_{2} \\
-\Phi A_{\rm\bf zx}^{[0]} & -\Phi F_{4} & I_{M_{\rm\bf v}}\!-\!\Phi A_{\rm\bf zv}^{[0]} & -\Phi J_{3} \end{array}\!\!\right]\!
\left[\!\!\begin{array}{c} \alpha \\ \xi \\ \beta \\ \eta \end{array}\!\!\right]\!=\!0
\label{eqn:a13}
\end{equation}

Combining Equations (\ref{eqn:a11})-(\ref{eqn:a13}) together, the following equality is obtained,
\begin{equation}
\left[\!\!\!\begin{array}{cccc}
\lambda_{0} E^{[0]}\!-\!A_{\rm\bf xx}^{[0]} & \lambda_{0} F_{1}\!-\!F_{2} & -A_{\rm\bf xv}^{[0]} & -J_{1} \\
-C_{\rm\bf x}^{[0]} & -F_{3} & -C_{\rm\bf v}^{[0]}  & -J_{2} \\
-P_{1}G & M\!-\!P_{1}H & 0 & 0 \\
0 & 0 & -P_{2}K & N\!-\!P_{2}S \\
-\Phi A_{\rm\bf zx}^{[0]} & -\Phi F_{4} & I_{M_{\rm\bf v}}\!-\!\Phi A_{\rm\bf zv}^{[0]} & -\Phi J_{3} \end{array}\!\!\right]\!\left[\!\!\begin{array}{c} \alpha \\ \xi \\ \beta \\ \eta \end{array}\!\!\!\!\right]\!\!=\!0
\label{eqn:a14}
\end{equation}

Obviously, the vector ${\rm\bf col}\{\alpha, \; \xi, \; \beta, \; \eta\}$ is not a zero vector. The definition of the matrix pencil ${\Xi}^{[o]}_{p}(\lambda)$ implies that the matrix ${\Xi}^{[o]}_{p}(\lambda_{0})$ is not of FCR.

On the contrary, assume that the matrix pencil ${\Xi}^{[o]}_{p}(\lambda)$ is rank deficient at a particular complex $\lambda_{0}$. Similar arguments show that at this $\lambda_{0}$, the matrix pencil ${\Xi}^{[o]}(\lambda)$ is also rank deficient.

This completes the proof.   \hspace{\fill}$\Diamond$

\hspace*{-0.45cm}{\rm\bf Proof of Corollary 2:} It can be declared from Lemmas \ref{lemma:1} and \ref{lemma:3} that, the number $\zeta_{Li}$ of Equation (\ref{eqn:17}) is equal to zero for each $i=1,2,\cdots,N$, if and only if there exists a $\lambda_{0}\in{\cal C}$, such that the matrix pencil $\bar{\Xi}^{[o]}(\lambda)$ is of FCR at this particular value. Here, the matrix pencil $\bar{\Xi}^{[o]}(\lambda)$ is defined as
\begin{equation}
\hspace*{-0.5cm} \bar{\Xi}^{[o]}(\lambda)=\left[\begin{array}{cc}
\lambda E(p)-A_{\rm\bf xx}(p) & -A_{\rm\bf xv}(p) \\
-C_{\rm\bf x}(p) & -C_{\rm\bf v}(p) \end{array}\right]
\label{eqn:a21}    \\
\end{equation}
That is, the matrix pencil $\bar{\Xi}^{[o]}(\lambda)$ is of FNCR.

Let $\tilde{\Xi}^{[o]}(\lambda_{0})$ represent the following matrix
\begin{equation}
\hspace*{-0.0cm} \left[\!\!\!\!\begin{array}{c}
{\rm\bf diag}\!\left\{\!\left.\Psi^{[0]}(\lambda_{0},i)\right|_{i=1}^{N}\!\!\right\} \\
\left\{M(i)N_{\rm\bf f}(i)\!-\!P_{1}\left[\!G(i)N^{[0]}_{\rm\bf x}(i)\!+\!H(i)N_{\rm\bf f}(i)\right]\right\}\!V_{i,0}^{-1}(m^{[0]}(i))   \\
\left\{N(i)N_{\rm\bf j}(i)\!-\!P_{2}\left[\!K(i)N^{[0]}_{\rm\bf v}(i)\!+\!S(i)N_{\rm\bf j}(i)\right]\right\}\!V_{i,0}^{-1}(m^{[0]}(i))
\end{array}\!\!\!\!\right]
\label{eqn:a22}
\end{equation}
in which $\Psi^{[0]}(\lambda_{0},i)$ with $1\leq i\leq N$ stands for the matrix
${\rm\bf diag}\!\left\{\!H_{\xi^{[0]}_{Hi}}(\lambda_{0}),\;K_{\xi^{[0]}_{Ki}(j)}(\lambda_{0})|_{j=1}^{\zeta^{[0]}_{Ki}},\; L_{\xi^{[0]}_{Li}(j)}(\lambda_{0})|_{j=1}^{\zeta^{[0]}_{Li}}\!\right\}$. On the basis of the above observations, similar arguments as those in the proofs of Theorems \ref{theorem:4} and \ref{theorem:5} show that, the matrix $\bar{\Xi}^{[o]}(\lambda_{0})$ is of FCR, if and only if  the matrix $\tilde{\Xi}^{[o]}(\lambda_{0})$ satisfies this requirement.

Assume now that $\zeta^{[0]}_{Li}=0$ for each $i=1,2,\cdots,N$. Then according to Lemma \ref{lemma:0}, the set ${\rm\bf\Lambda}^{[0]}$ consists of only finitely many elements. This means that the set ${\cal C}\backslash {\rm\bf\Lambda}^{[0]}$ is not empty. Moreover, for each $\lambda_{0}\in {\cal C}\backslash {\rm\bf\Lambda}^{[0]}$ and each $i=1,2,\cdots,N$, the matrix $\Psi^{[0]}(\lambda_{0},i)$ is of FCR, which further leads to that the matrix ${\rm\bf diag}\!\left\{\!\left.\Psi^{[0]}(\lambda_{0},i)\right|_{i=1}^{N}\!\!\right\}$ is of FCR. It can therefore be claimed from Equation (\ref{eqn:a22}) that the matrix $\tilde{\Xi}^{[o]}(\lambda_{0})$ is of FCR. Hence, the matrix pencil $\bar{\Xi}^{[o]}(\lambda)$ is of FNCR, which is equivalent to $\zeta_{Li}=0$ for each $1\leq i\leq N$.

Now, assume that there exists a $i\in \{1,2,\cdots,N\}$, such that $\zeta^{[0]}_{Li}>0$. Then according to Lemma \ref{lemma:0}, the set ${\rm\bf\Lambda}^{[0]}$ is equal to the whole complex plane. That is, ${\rm\bf\Lambda}^{[0]}={\cal C}$.

If there exists a $\lambda_{0}\in {\rm\bf\Lambda}_{o}$, such that for each $i=1,2,\cdots,N$, the matrix of Equation (\ref{eqn:23}) is of FCR, then from the definition of the matrices $\bar{N}^{[0]}(\lambda_{0},i)|_{i=1}^{N}$ and Lemma \ref{lemma:3}, it can be claimed that the matrix $\tilde{\Xi}^{[o]}(\lambda_{0})$, and hence the matrix $\bar{\Xi}^{[o]}(\lambda_{0})$, is of FCR. Therefore, $\zeta_{Li}=0$ for every $1\leq i\leq N$.

On the contrary, assume that for each $\lambda_{0}\in {\rm\bf\Lambda}_{o}$, there is at least one $i\in \{1,2,\cdots,N\}$, such that the matrix of Equation (\ref{eqn:23}) is not of FCR. Then the above arguments imply that for every $\lambda_{0}\in {\rm\bf\Lambda}_{o}$, the matrix $\bar{\Xi}^{[o]}(\lambda_{0})$ is not of FCR. Hence, the determinant of every $(M_{\rm\bf x}+M_{\rm\bf v})\times (M_{\rm\bf x}+M_{\rm\bf v})$ dimensional submatrix of the matrix $\bar{\Xi}^{[o]}(\lambda_{0})$ is equal to zero.

Note that each $(M_{\rm\bf x}+M_{\rm\bf v})\times (M_{\rm\bf x}+M_{\rm\bf v})$ dimensional submatrix of the matrix pencil $\bar{\Xi}^{[o]}(\lambda)$ is still a matrix pencil. Moreover, its determinant is a polynomial of the variable $\lambda$ with a degree at most $M_{\rm\bf x}+M_{\rm\bf v}$. On the other hand, the set ${\rm\bf\Lambda}_{o}$ consists of $M_{\rm\bf x}+M_{\rm\bf v}+1$ different elements. It can therefore be declared that if the 2nd condition of this corollary is not satisfied, then the determinant of every $(M_{\rm\bf x}+M_{\rm\bf v})\times (M_{\rm\bf x}+M_{\rm\bf v})$ dimensional submatrix of the matrix pencil $\bar{\Xi}^{[o]}(\lambda)$ is a zero polynomial. Hence, it is rank deficient at each $\lambda\in{\cal C}$. That is, the matrix pencil $\bar{\Xi}^{[o]}(\lambda)$ is not of FCR at every $\lambda\in{\cal C}$, which is equivalent to that this matrix pencil is not of FNCR.

This completes the proof.   \hspace{\fill}$\Diamond$

\end{document}